\documentclass[fleqn,usenatbib]{mnras}

\usepackage[T1]{fontenc}

\DeclareRobustCommand{\VAN}[3]{#2}
\let\VANthebibliography\thebibliography
\def\thebibliography{\DeclareRobustCommand{\VAN}[3]{##3}\VANthebibliography}

\usepackage{graphicx}	
\usepackage{amsmath}	
\usepackage{amssymb}	
\usepackage{mathtools}
\usepackage{hyperref}
\usepackage{natbib}
\usepackage{caption}
\usepackage{subcaption}
\usepackage{gensymb}
\usepackage[shortlabels]{enumitem}
\usepackage{newtxtext,newtxmath}
\usepackage{booktabs}
\usepackage{appendix}

\newcommand{\abssize}{\vert \Delta \Omega_{\rm C} \vert / \Omega_0}
\newcommand{\deltat}{\Delta t / T_0}
\newcommand{\vstress}{v_{\rm stress}}
\newcommand{\OmegaC}{\Omega_{\rm C}}

\title[Anti-glitches from vortex avalanches]{Anti-glitches in accreting pulsars from superfluid vortex avalanches}

\author[G. Howitt et al.]{
G. Howitt,$^{1,2}$\thanks{E-mail: gawhowitt@gmail.com}
A. Melatos,$^{1,2}$
\\
$^{1}$School of Physics, University of Melbourne, Parkville, Victoria, 3010 Australia\\
$^{2}$OzGrav, Australian Research Council Centre of Excellence for Gravitational Wave Discovery, University of Melbourne, Victoria, 3010 Australia\\
$^{3}$Peter MacCallum Cancer Centre, Melbourne, Victoria, 3000 Australia
}	

\date{Accepted XXX. Received YYY; in original form ZZZ}

\pubyear{2015}

\begin{document}
\label{firstpage}
\pagerange{\pageref{firstpage}--\pageref{lastpage}}
\maketitle

\begin{abstract}
Three sudden spin-down events, termed `anti-glitches', were recently discovered in the accreting pulsar NGC 300 ULX-1 by the \textit{Neutron Star Interior Composition Explorer} (NICER) mission.
Unlike previous anti-glitches detected in decelerating magnetars, these are the first anti-glitches recorded in an accelerating pulsar.
One standard theory is that pulsar spin-up glitches are caused by avalanches of collectively unpinning vortices that transfer angular momentum from the superfluid interior to the crust of a neutron star. 
Here we test whether vortex avalanches are also consistent with the anti-glitches in NGC 300 ULX-1, with the angular momentum transfer reversed. 
We perform $N$-body simulations of up to $5 \times 10^{3}$ pinned vortices in two dimensions in secularly accelerating and decelerating containers.
Vortex avalanches routinely occur in both scenarios, propagating inwards and outwards respectively. 
The implications for observables, such as size and waiting time statistics, are considered briefly.

\end{abstract}




\section{Introduction}

Around 10\% of known pulsars experience sudden changes in spin frequency known as glitches.
Typically, glitches occur in isolated pulsars and are observed as a near-instantaneous increase $\Delta \nu$ in frequency $\nu$ that interrupts the stars' steady spin down caused by electromagnetic braking
\citep{Lyne2006}. 
Glitches occur at random times with intervals from weeks to years and with fractional sizes spanning $10^{-10} \lesssim \Delta \nu / \nu \lesssim 10^{-5}$.
Most glitches are discovered through a secular drift in the phase residuals of pulsar timing experiments, which can be undone by including a step change in the frequency $\nu$, often followed by a transient increase in the pulsar's spin-down rate $\dot{\nu}$
\citep{Wang2000}.
The epoch and size of the glitch are determined by least-squares fitting to a phase-connected timing model between the observations.
Other detection techniques also exist, such as the hidden Markov scheme described in 
\citet{Melatos2020}.
\citet{Palfreyman2018} 
observed a glitch in the Vela pulsar in real time, and 
\citet{Ashton2019} used Bayesian parameter estimation to track the frequency evolution during the glitch.
Glitches do not occur in isolated electromagnetically braking pulsars exclusively.
Spin-up glitches have also been observed in 
accretion-powered pulsars 
\citep{Serim2017} 
and magnetars
\citep{Dib2008}.

Anti-glitches, i.e. sudden \textit{decreases} in the spin frequency, are rarer than spin-up glitches.
They have been observed in accretion-powered pulsars 
\citep{Ray2019}
and magnetars 
\citep{Dib2008}.
The situation in magnetars is complicated: anti-glitches and spin-up glitches can occur in the same object, e.g. 1E 2259+586
\citep{Kaspi2003,Archibald2013,Younes2020},
they are accompanied sometimes by radiative changes in the pulse profile and X-ray flux
\citep{Dib2008,Dib2014}, 
and they may be triggered or modified by internal reorganization of the magnetic field instead of electromagnetic braking
\citep{Garcia2015,Mastrano2015}.
We therefore elect not to model magnetar anti-glitches in this paper. 
In contrast, anti-glitches in accretion-powered pulsars present a simpler physical phenomenon.
They are essentially the reverse of glitches in rotation-powered pulsars --- that is, impulsive spin-down events which occur when the star accelerates secularly under the action of an accretion torque.
They are the focus of the present work. 

Recent observations from the \textit{Neutron Star Interior Composition Explorer} (NICER) mission on the International Space Station 
\citep{Gendreau2016}
reported three spin-down glitches in the accreting pulsar NGC 300 ULX-1
\citep{Ray2019}.
Ultraluminous X-ray sources (ULXs) are extragalactic X-ray-bright point sources with super-Eddington luminosities, some of which (though not all) host neutron stars
\citep{King2016,Kaaret2017}.
NGC 300 ULX-1 is a high-mass X-ray binary in the nearby galaxy NGC 300, which is visible only in outburst
\citep{Monard2010,Binder2011}.
\citet{Carpano2018} detected pulsations in NGC 300 ULX-1, unambiguously identifying it as a neutron star, and a NICER campaign in 2018 timed this object daily for $\approx 100$ d before it became too faint.
During this interval, \citet{Ray2019} constructed a coherent timing model over several contiguous data segments of $\approx 10$ d each, measuring $\dot{\nu} = 4.3 \times 10^{-10}$ Hz s$^{-1}$ (spin up).
\citet{Ray2019} 
also measured an anti-glitch with $\Delta \nu / \nu = -4.4 \times 10^{-4}$ at MJD 58243 and a second anti-glitch with $\Delta \nu / \nu = -5.5 \times 10^{-4}$ at MJD 58265.
A third anti-glitch with $\Delta \nu / \nu = -7 \times 10^{-4}$ occurred at MJD 58334, however a phase-coherent timing model could not be constructed around this event as the X-ray flux dropped. 
No spectral changes were detected around any of the glitches, however, this finding is not conclusive due to the inherent variability of NGC 300 ULX-1.
Unlike the anti-glitching magnetar 1E 2259+586, NGC 300 ULX-1 has $\dot{\nu} > 0$ due to accretion.	
It is worth noting that not all accreting pulsars spin up
\citep{Bildsten1997,Long2005}.
Another ULX, M82-X2, experienced a spin-up glitch 
\citep{Bachetti2020}.
Interestingly, this glitch occurred while M82-X2 was observed to be spinning down, despite previous observations measuring it to be spinning up 
\citep{Bachetti2014}.

A popular theoretical explanation of glitches is that they are caused by superfluid vortex avalanches 
\citep{Anderson1975,Warszawski2011,Haskell2015},
although alternatives such as starquakes are also viable
\citep{Ruderman1976,Alpar1996,Middleditch2006,Chugunov2010,Giliberti2021}.
Quantized vortices in the superfluid inner crust are postulated to be pinned metastably at nuclear lattice sites or magnetic flux tubes locked to the crust, until the spin-down-driven crust-superfluid lag exceeds a threshold, and the vortices unpin collectively via knock-on processes
\citep{Warszawski2011}
and migrate radially outwards en masse, transferring angular momentum from the superfluid to the crust 
\citep{Anderson1975}. 
This mechanism has been explored extensively for spin-up glitches 
\citep{Haskell2015}
but there is every reason a priori to think that it applies in reverse to anti-glitches too, such as those in accreting pulsars, with vortices migrating inwards instead of outwards 
\citep{Ray2019}. 

In this paper we test this idea rigorously, by solving the equations of motion self-consistently for $\sim 10^4$ vortices using an $N$-body solver calibrated against traditional spin-up glitches in a decelerating container 
\citep{Howitt2020}. 
The simulations demonstrate that spinning up the container results in avalanches where vortices move inward, causing anti-glitches. 
The paper is organized as follows.
In section \ref{sec:vortex dynamics}, we provide some background on the vortex avalanche model for spin-up and spin-down glitches.
In section \ref{sec:method},
we describe the mathematical framework of the $N$-body point vortex model and its equations of motion.
We also outline the numerical solver, including a new algorithm for adding additional vortices as the angular velocity of the container increases.
In section \ref{sec:results} we present the results of simulations of vortex avalanches leading to anti-glitches and compare the results with traditional spin-up glitches, when the torque on the container reverses sign.
In section \ref{sec:idealizations}, we discuss idealizations inherent to the model.
In section \ref{sec:discussion}, we apply the results to observations of glitches and the anti-glitches in NGC 300 ULX-1 and contrast its behaviour with that observed in magnetars.

\section{Superfluid vortex avalanches}
\label{sec:vortex dynamics}

To orient the reader, we begin with a brief review of the standard picture of vortex avalanches for traditional spin-up glitches in section \ref{subsec:glitch background}.
We then explain qualitatively in section \ref{subsec:antiglitch background} how the same mechanism is expected to operate in reverse for anti-glitches and foreshadow some subtle differences between the two ``directions'' of vortex migration. 
These qualitative descriptions are then tested quantitatively with the $N$-body solver 
\citep{Howitt2020}
in sections \ref{sec:method} onwards.

\subsection{Spin-up glitches in isolated pulsars}
\label{subsec:glitch background}

In a superfluid, rotation is supported by the formation of an array of vortices with quantized circulation
\citep{Onsager1949,Feynman1955}.
The inner crust of a neutron star is expected to contain $\sim 10^{15}$ vortices
\citep{Baym1969a,Alpar1977}.
At length scales much greater than the inter-vortex separation, the rotational characteristics of a superfluid mimic solid body rotation
\citep{Osborne1950}.
For a container rotating with angular velocity $\Omega$, the number of vortices $N_{\rm v}$ within radius $R$ is determined by the Feynman condition 
\citep{Feynman1955}
\begin{equation}
N_{\rm v} \kappa = \Omega R^2 \, ,
\label{eq:feynman condition}
\end{equation}
where $2 \pi \kappa$ is the quantum of circulation.

Equation \eqref{eq:feynman condition} implies that vortices move outward as the star spins down due to electromagnetic braking.
However, the outward motion is frustrated by nuclear lattice sites or magnetic flux tubes, which pin vortices, leading to differential rotation between the viscous component of the star (which is coupled to the crust) and the inviscid superfluid component
\citep{Alpar1977}.
The lag between the components builds until a (local) threshold is exceeded and individual vortices unpin.
The unpinned vortices trigger knock-on unpinning of nearby vortices, e.g. via proximity or acoustic effects, leading to a vortex avalanche
\citep{Warszawski2012a}.
The motion of the vortices transfers angular momentum from the superfluid to the crust or vice versa, depending on the direction in which they move.
If the vortices move radially outward on average, the superfluid spins down, and the crust spins up
\citep{Pethick2001}.

\subsection{Anti-glitches in accreting pulsars}
\label{subsec:antiglitch background}

Equation \eqref{eq:feynman condition} also implies that if the star spins up (e.g. by accretion), then vortices tend to migrate inward, as $N_{\rm v}/R^2$ increases.
Under these circumstances, the scenario in section \ref{subsec:glitch background} runs in reverse.
Pinning at nuclear sites or magnetic flux tubes frustrates the inward migration; vortices unpin collectively in an avalanche mediated by knock-on processes, when the crust-superfluid lag exceeds a threshold; vortices move radially inward during the avalanche; and the angular momentum of the superfluid increases, while the angular momentum of the crust decreases. 
The result is an anti-glitch with $\Delta \nu < 0$.
The process by which vortices nucleate at the outer boundary and enter the superfluid, as the superfluid spins up, is approximately but not exactly a reversal of the process by which they exit the superfluid, as the superfluid spins down
\citep{Stagg2016}.

\section{Point Vortex Dynamics}
\label{sec:method}

When the inter-vortex separation is much greater than the vortex core radius, the quantum mechanical structure of the vortex cores can be ignored and vortices are treated as point-like filamentary features in the velocity field obeying classical hydrodynamics
\citep{Bustamante2015}. 
In the absence of external forces such as pinning or dissipation, the velocity d$\mathbf{x}$/d$t$ of a vortex at $\mathbf{x}(t)$ is equal to the bulk fluid velocity at $\mathbf{x}(t)$ induced by the other vortices, which can be computed directly from the vorticity distribution.
In the system studied here, the vortices are also influenced by interactions with impurities, the container boundary and a viscous fluid component which co-rotates with the container. 

In this paper, we consider an array of point vortices in a circular container with cylindrical symmetry.
The restriction to two dimensions assumes that the vortex filaments are infinitely long, rigid and aligned parallel to the rotation axis.
This assumption ignores the potentially important three-dimensional dynamics of vortex tangles and vortex tension
\citep{Peralta2006b, Mongiovi2017,  Drummond2017, Drummond2018, Haskell2020}
but is a necessary simplification in order to simulate $N_{\rm v}  \gtrsim 10^3$ vortices in order for collective avalanche motion to be observable.

\subsection{Equations of motion}
\label{subsec:equations of motion}

In a reference frame rotating with angular velocity $\Omega$, the position of a vortex at Cartesian coordinates, $(x_i,$ $y_i)$, evolves according to
\begin{equation}
\frac{d}{dt} 
\begin{pmatrix}
x_i \\
y_i
\end{pmatrix}
= \mathcal{R}_\phi
\begin{pmatrix}
v_{i,x} \\
v_{i,y}
\end{pmatrix} \, ,
\label{eq:total velocity}
\end{equation}
with
\begin{equation}
v_{i,x} = - \sum_{j \neq i} \frac{\kappa y_{ij}}{r_{ij}^2} 
+ \sum_{j=1}^{N_{\rm v}} \frac{\kappa y_{ij,\rm{image}}}{r_{ij,\rm{image}}^2}
+ \Omega y_i
- \sum_k \frac{\partial V(\textbf{x}_i-\textbf{x}_k)}{\partial y_i} 
\label{eq:x velocity}
\end{equation}
\begin{equation}
v_{i,y} = \sum_{j \neq i} \frac{\kappa x_{ij}}{r_{ij}^2} 
- \sum_{j=1}^{N_{\rm v}} \frac{\kappa x_{ij,\rm{image}}}{r_{ij,\rm{image}}^2}
- \Omega x_i
+ \sum_k \frac{\partial V(\textbf{x}_i - \textbf{x}_k)}{\partial x_i} \, .
\label{eq:y velocity}
\end{equation}
In equations \eqref{eq:x velocity} and \eqref{eq:y velocity},
$2 \pi \kappa$ is the quantum of circulation and we define 
$\textbf{x}_{ij} = \textbf{x}_i - \textbf{x}_j = (x_{ij},y_{ij})$ to be the displacement between vortices at $\textbf{x}_i$ and $\textbf{x}_j$, with $r_{ij} = \vert \textbf{x}_{ij} \vert$.
We also define $\textbf{x}_{ij, \rm{image}} = \textbf{x}_i - \textbf{x}_{j, \rm{image}} = (x_{ij,\rm{image}},y_{ij,\rm{image}})$ to be the displacement between a vortex at $\textbf{x}_i$ and the image vortex of a vortex at $\textbf{x}_j$, with 
$r_{ij,{\rm image}} = \vert \textbf{x}_{ij,{\rm image}} \vert$.
Image vortices enforce a no-penetration boundary condition of the superfluid at the boundary in the dissipation-free regime
\citep{Schwarz1985}.
The function 
$V(\mathbf{x} - \mathbf{x}_k) = -V_0 \exp[-(\mathbf{x}-\mathbf{x}_k)^2/2 \xi^2]$ describes the attractive pinning potential with depth $V_0$ and characteristic width $\xi$ due to a pinning site at $\mathbf{x}_k$.
In equation \eqref{eq:total velocity}, the term $\mathcal{R}_\phi$ describes rotation of the velocity vector through a `dissipation angle' $\phi$, whose value relates to the strength of interaction between the inviscid and viscous components of the superfluid 
\citep{Campbell1979,Sedrakian1995}.
We use a dimensionless coordinate system, where $\kappa$ and the fundamental length unit both equal unity. 
All other quantities, such as time and velocity, are defined through these quantities.
A more detailed explanation of equations \eqref{eq:total velocity}--\eqref{eq:y velocity} is provided in \citet{Howitt2020}.

\subsection{$N$-Body Solver}
\label{subsec:numerical method}

In this paper, we use the same Python code for solving equations \eqref{eq:total velocity}--\eqref{eq:y velocity} described in 
\citet{Howitt2020}.

Simulations are initialized by creating an ensemble of $N_{\rm v}$ point vortices within a circular container of radius $R$.
The initial positions are drawn at random from a uniform spatial distribution.
The container contains a square grid of pinning sites with grid spacing $a$, pinning depth $V_0$ and characteristic width $\xi$.
The initial angular velocity of the container $\Omega(t=0)$ is determined from equation \eqref{eq:feynman condition}.
At each time step, we compute the velocity of all vortices from equations \eqref{eq:total velocity}--\eqref{eq:y velocity} and step forward in time using an adaptive time-step Runge-Kutta-Cash-Karp (RKCK) scheme. 
If any vortices leave the container, their circulation is set to zero and their contribution to equations \eqref{eq:total velocity}--\eqref{eq:y velocity} is ignored.

After each time step, we evolve the angular velocity of the container $\Omega_{\rm C}$ by solving 
\begin{equation}
	\frac{d \Omega_{\rm C}} {dt} = N_{\rm ext} - I_{\rm rel} \frac{d \Omega _{\rm S}} {dt} \quad .
	\label{eq:angular velocity evolution}
\end{equation}
Here $N_{\rm ext}$ is the external torque divided by the container moment of inertia $I_{\rm C}$, and we set $I_{\rm rel} = I_{\rm S} / I_{\rm C} = 1$ in all simulations described in this paper, where $I_{\rm S}$ is the moment of inertia of the superfluid treated notionally as a rigid body.
The angular velocity of the superfluid $\Omega_{\rm S}$ is determined by the configuration of the vortex array
\citep{Fetter1966, Howitt2020}.
We assume the change in angular momentum of the vortex array is transferred conservatively to the container
\citep{Pethick2001}. 

In order to test whether the vortex code can produce anti-glitches similar to those in NGC 300 ULX-1, we perform simulations on pinned vortices in an accelerating container.
In this scenario, vortices tend to move toward the centre of the container in order to maintain the Feynman condition \eqref{eq:feynman condition}.
Therefore, as the container spins up, new vortices must be added, otherwise an annular region around the boundary becomes depleted (unphysically) of vortices.
This issue does not arise in 
\citet{Howitt2020},
where only decelerating containers are considered.
In terrestrial superfluids, it is observed that vortices enter the system at the outer boundary, where the density is lowest, and vortex nucleation is favoured energetically
\citep{Donnelly1991,Stagg2016}.
For this paper, we modify the code to include a simple algorithm which inserts a new vortex at $r = 0.999R$ and at a random angular coordinate, whenever $\Omega$ increases by $\kappa / R^2$ since the last vortex was added.

\section{Anti-glitch simulations}
\label{sec:results}

We perform a suite of numerical experiments to investigate whether anti-glitches in an accelerating container are simply the time-reversed analogue of glitches during spin down.
The system parameters for the simulations are, cf. section 5.1 of 
\citet{Howitt2020}: 
$R=10$, $\Delta t = 0.1 T_0$ (where $T_0 = 2 \pi R^2 / N_{\rm v} \kappa$ is the initial rotation period of the container), $I_{\rm rel} = 1$, $V_0=2000$, $a=10^{-2}R$ (i.e. there are approximately 10 pinning sites for each vortex initially in the simulation), $\xi = 10^{-3} R$, $\phi = 0.1$ rad.
Here $I_{\rm rel}$ is the moment of inertia of the container divided by the nominal moment of inertia of the superfluid, approximated for this purpose as a uniformly rotating rigid body, as in the standard two-component model of a neutron star
\citep{Baym1969a}.

We initialise the vortices by drawing their positions from a random spatial distribution.
We then evolve the system without spin up/down or container feedback until all vortices are pinned, which we define as when no vortices move a distance more than the pinning-site separation $a$ between successive time steps.

We perform four numerical experiments, described below and summarised in Table \ref{tab:numerical experiments}.
\begin{enumerate}[label=\Roman*]
	\item Spin up from the pinned initial state with $N_{\rm ext} = 10^{-3} \Omega_0 / T_0$ for $2 \times 10^5$ time steps, or $0 \leq t/T_0 \leq 2 \times 10^4$.
	\item Spin down from the pinned initial state with $N_{\rm ext} = -10^{-3} \Omega_0 / T_0$ for $2 \times 10^5$ time steps, or $0 \leq t/T_0 \leq 2 \times 10^4$.
	\item We resume experiment I where it left off at $t/T_0 = 2 \times 10^4$ and continue for $2 \times 10^4 < t/T_0 \leq 4 \times 10^4$ with $N_{\rm ext} = -10^{-3} \Omega_0 / T_0$, i.e. the sign of the torque is reversed.
	\item We resume experiment II where it left off at $t/T_0 = 2 \times 10^4$ and continue for $2 \times 10^4 < t/T_0 \leq 4 \times 10^4$ with $N_{\rm ext} = 10^{-3} \Omega_0 / T_0$.
\end{enumerate}
\begin{table*}
	\centering
	\begin{tabular}{|l|l|l|l|l|}
		& Experiment I & Experiment II & Experiment III & Experiment IV \\
		\hline
		$N_{v,i}$ & 1943 & 1943 & 2361  & 1498 \\ 
		$\Omega_i/\Omega_0$ & 1 & 1 & 1.27 & 0.74 \\
		$N_{v,f}$ & 2361 & 1498 & 1998 & 1888 \\
		$\Omega_f/\Omega_0$ & 1.27 & 0.74 & 0.99 & 1.01 \\
		$N_{\rm glitches}$ & 176 & 101 & 116 & 166 \\
		$N_{\rm ext}$ & $10^{-3} \Omega_0 / T_0$ & $-10^{-3} \Omega_0 / T_0$ & $-10^{-3} \Omega_0 / T_0$ & $10^{-3} \Omega_0 / T_0$ \\
		Time span & $0 \leq t/T_0 \leq 2 \times 10^4$ & $0 \leq t/T_0 \leq 2 \times 10^4$  & $2 \times 10^4 < t/T_0 \leq 4 \times 10^4$ & $2 \times 10^4 < t/T_0 \leq 4 \times 10^4$ \\
		\hline
	\end{tabular}
\caption{Summary of numerical experiments described in Section \ref{sec:results} showing initial and final numbers of vortices $N_{v,i}$ and $N_{v,f}$, initial and final container angular velocities $\Omega_i$ and $\Omega_f$, number of glitches $N_{\rm glitches}$, external torque $N_{\rm ext}$ and the time span of each simulation.}
\label{tab:numerical experiments}
\end{table*}
We use the same initial vortex configuration for both experiments I and II.
Experiments III and IV use the state of the vortex array at $t=2 \times 10^4 T_0$ in experiments I and II respectively as their initial condition. 

We describe the avalanche dynamics of anti-glitches in experiment I in section \ref{subsec:avalance dynamics}.
We compare the probability distribution functions of the glitch/anti-glitch size and waiting time probability density functions (PDFs) from experiments I and II in Section \ref{subsec:kdes}.
We find that the size and waiting time PDFs from experiments I and II are significantly different. 
We explore possible causes for this difference in Section \ref{subsec:differences} and in Appendix \ref{appendix}, which also includes discussion of experiments III and IV.
Finally, we vary the parameters $V_0$, $a$, $N_{\rm ext}$ and $\phi$ in the same way as in Section 5.4 of 
\citet{Howitt2020}
in order to explore how each parameter affects the properties of anti-glitches.
The results are presented in Section \ref{subsec:model variations}.

\subsection{Avalanche dynamics}
\label{subsec:avalance dynamics}

\begin{figure}
	\begin{center}
		\includegraphics[scale=0.6]{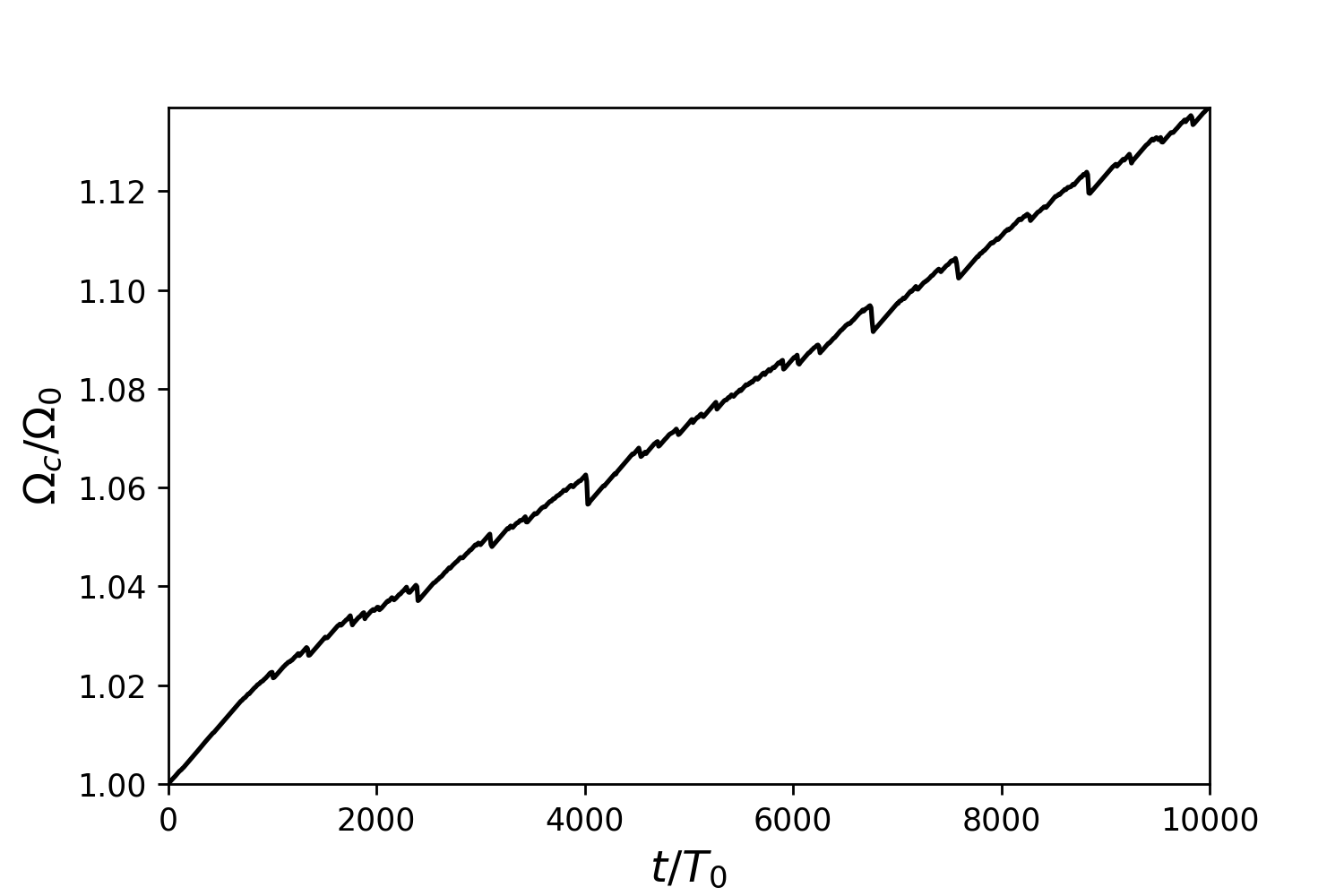} \\
		\includegraphics[scale=0.6]{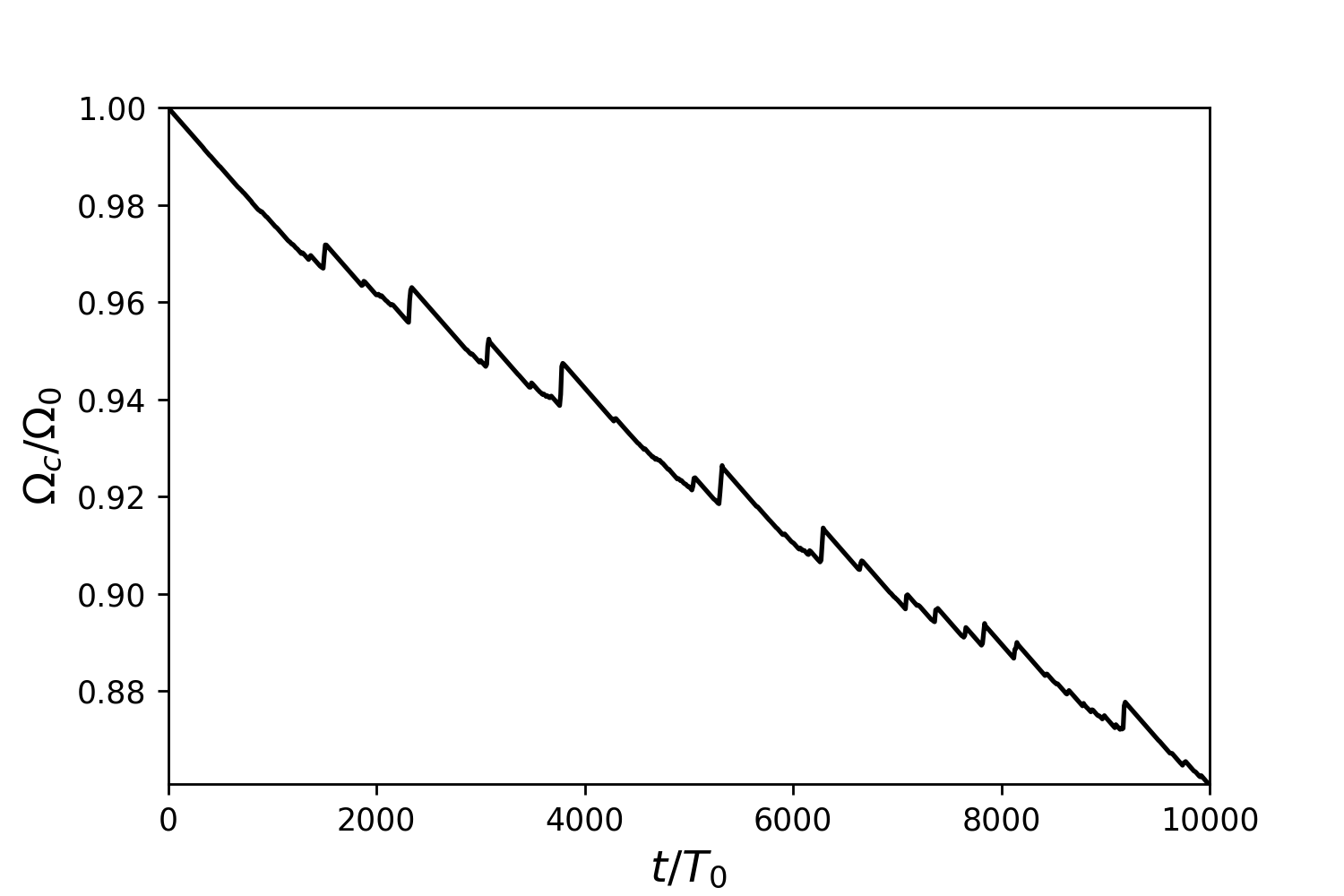}
		\caption{Container angular velocity $\OmegaC$, normalized by its initial value $\Omega_0$, as a function of time for $t < 10^4 T_0$.
		Top panel: Experiment I.
	Bottom panel: Experiment II.}
		\label{fig:spinup_frequency}
	\end{center}
\end{figure}

Figure \ref{fig:spinup_frequency} shows the evolution of the angular velocity of the container $\OmegaC$, normalized by its initial value $\Omega_0$ during experiment I (top panel) and experiment II (bottom panel).
For clarity, we show only $t < 10^4 T_0$. 
The behaviour of $\OmegaC$ in Figure \ref{fig:spinup_frequency} is qualitatively similar in both experiments I and II. 
The steady spin up/down is punctuated by abrupt spin down/up events of varying size at random times.

\begin{figure}
	\begin{center}
		\includegraphics[scale=0.19]{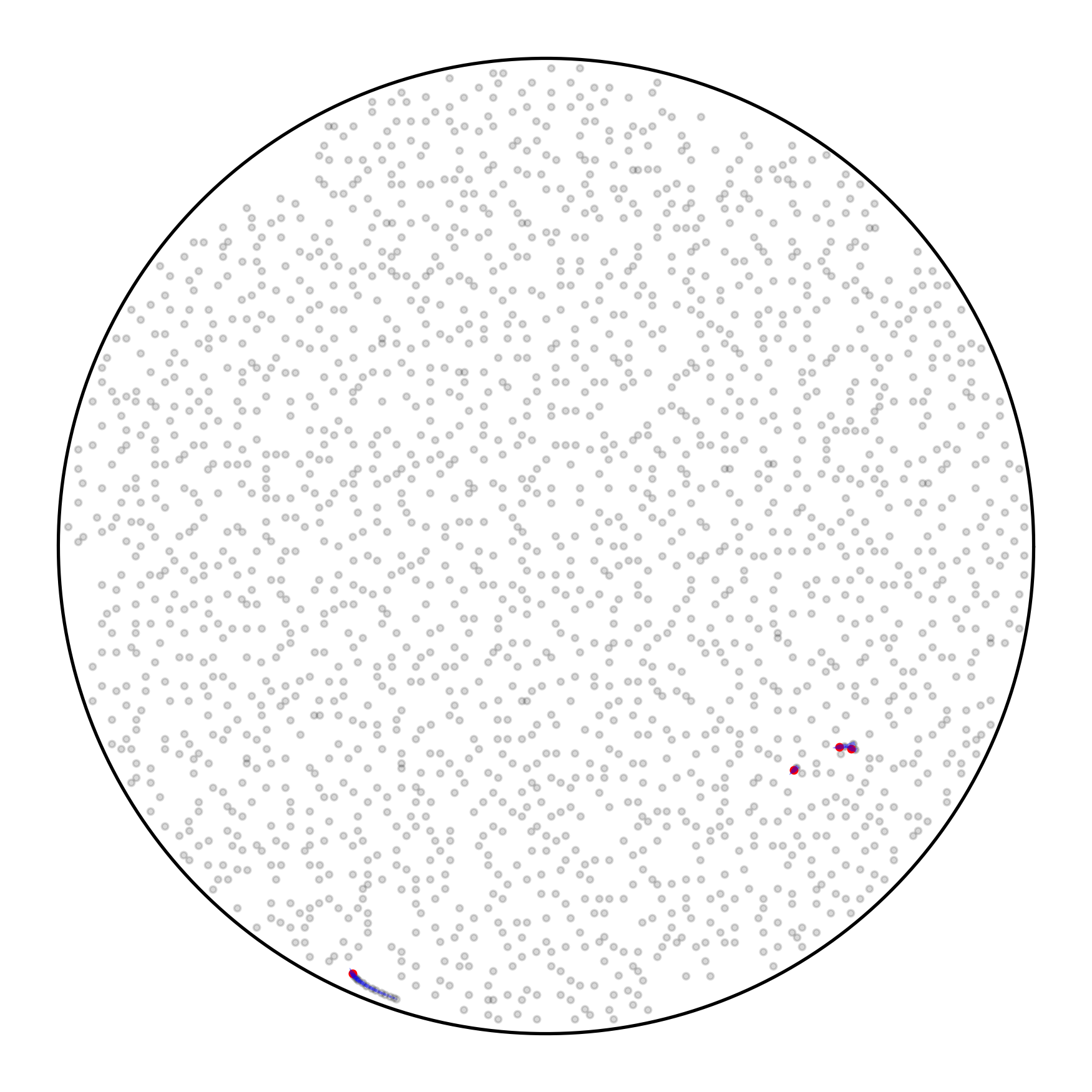}\\
		\includegraphics[scale=0.19]{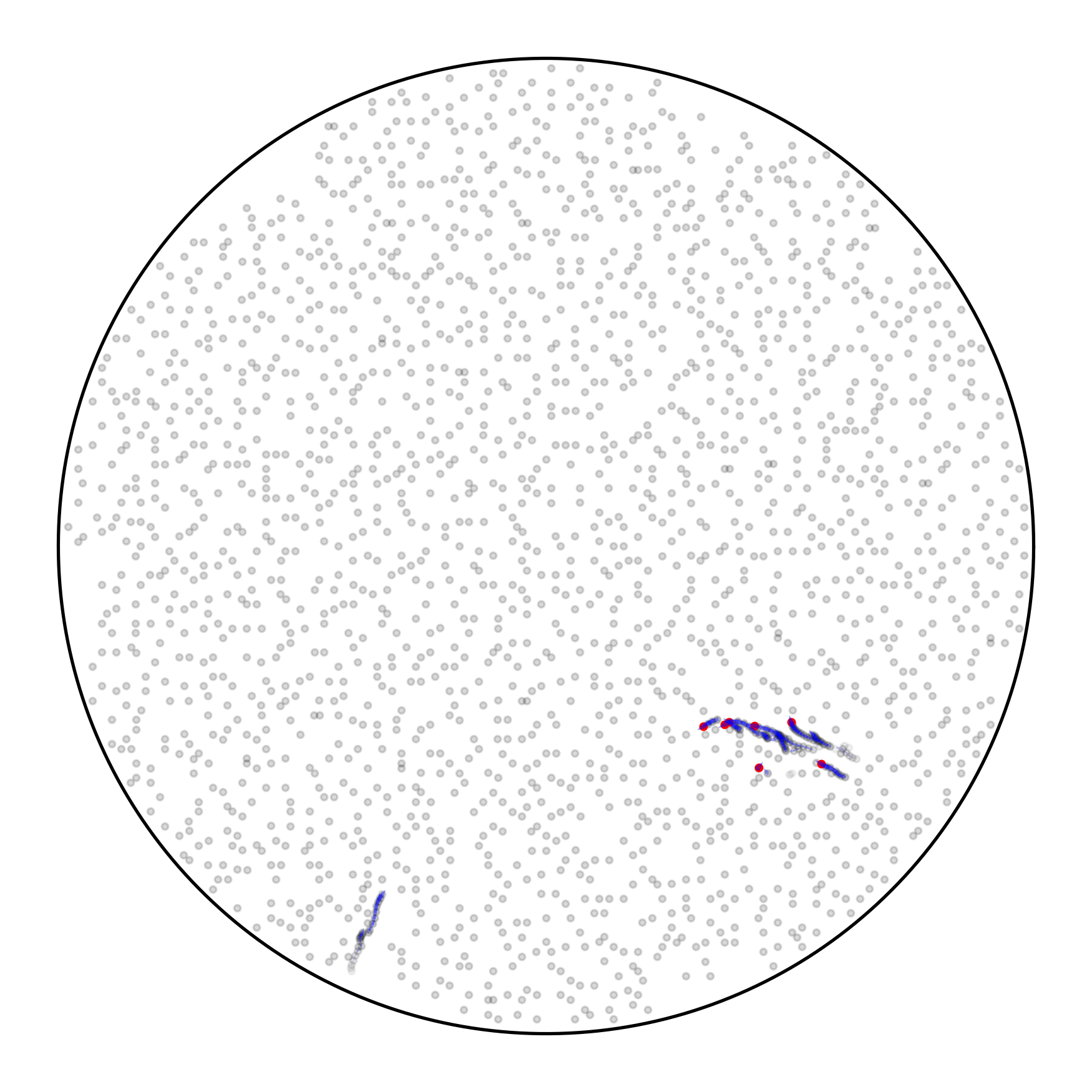}\\
		\includegraphics[scale=0.19]{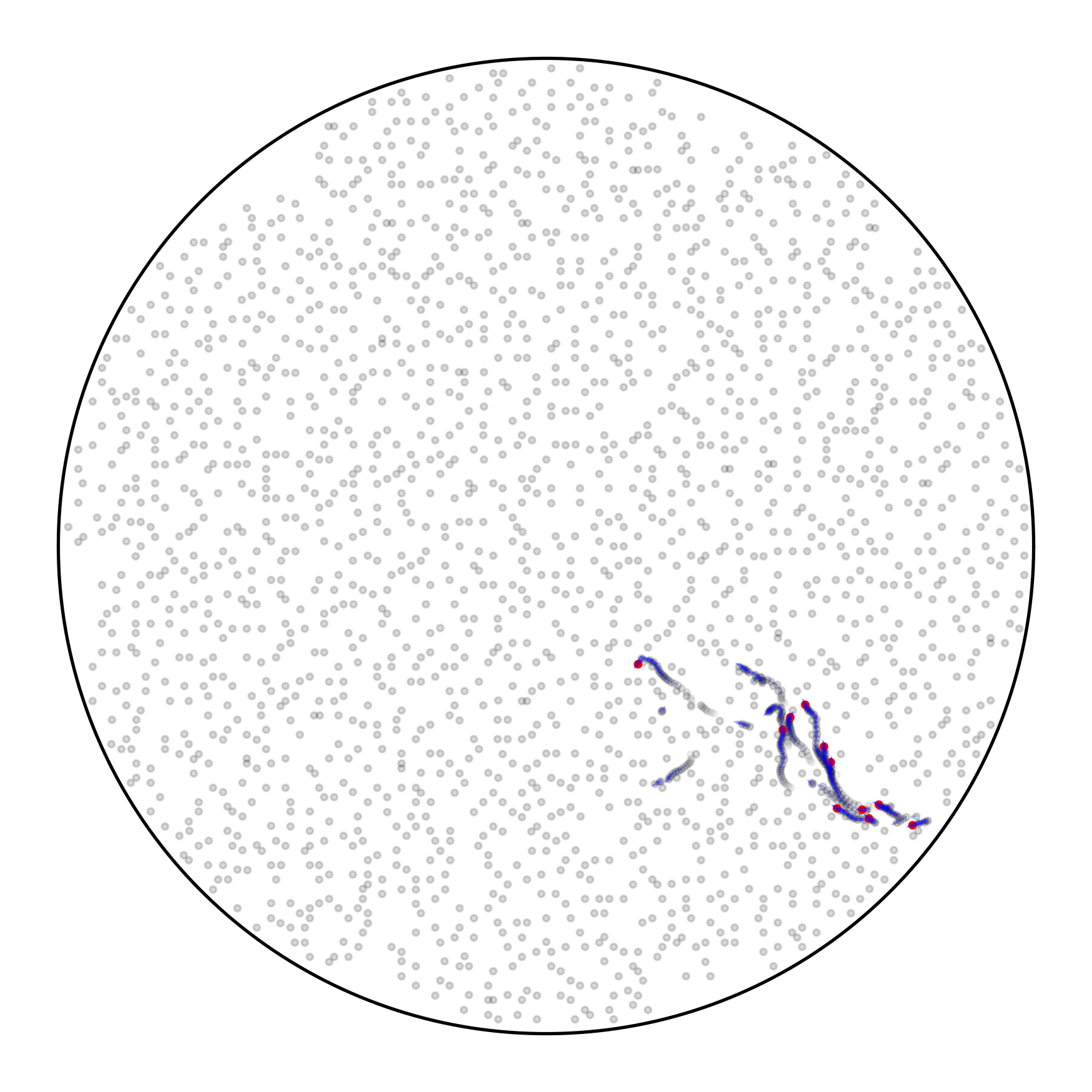}\\
		\includegraphics[scale=0.19]{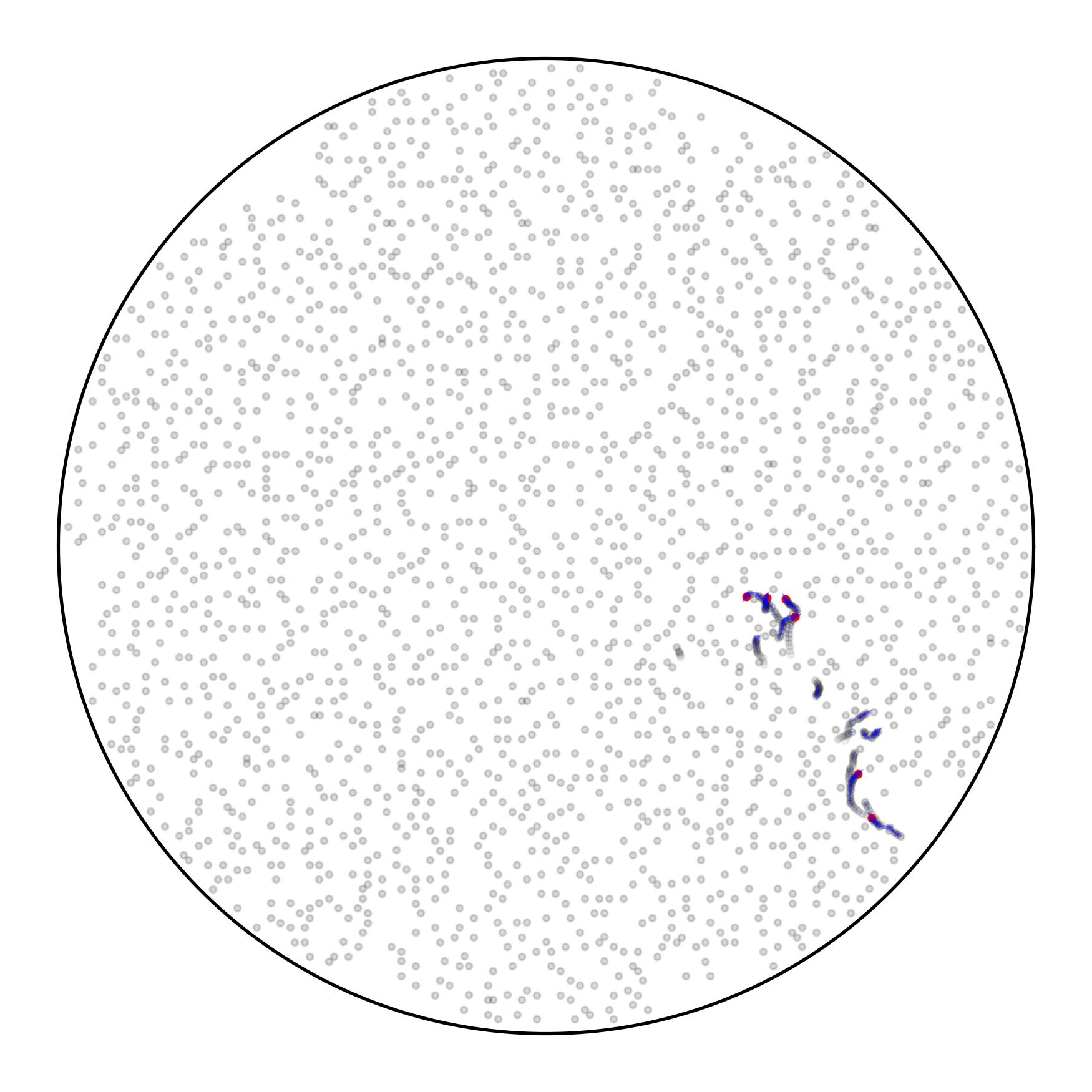}\\
		\includegraphics[scale=0.19]{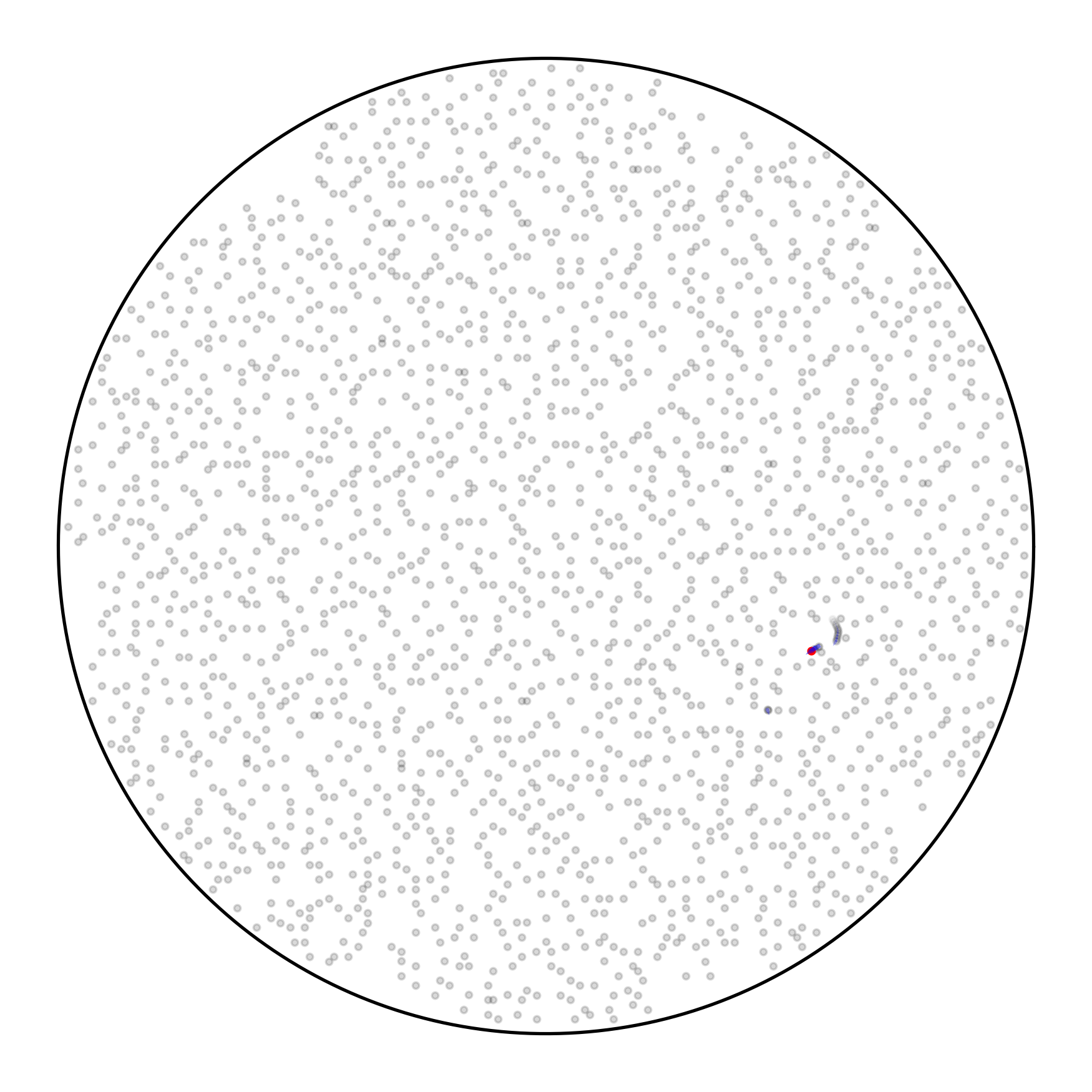}\\
		\includegraphics[scale=0.4]{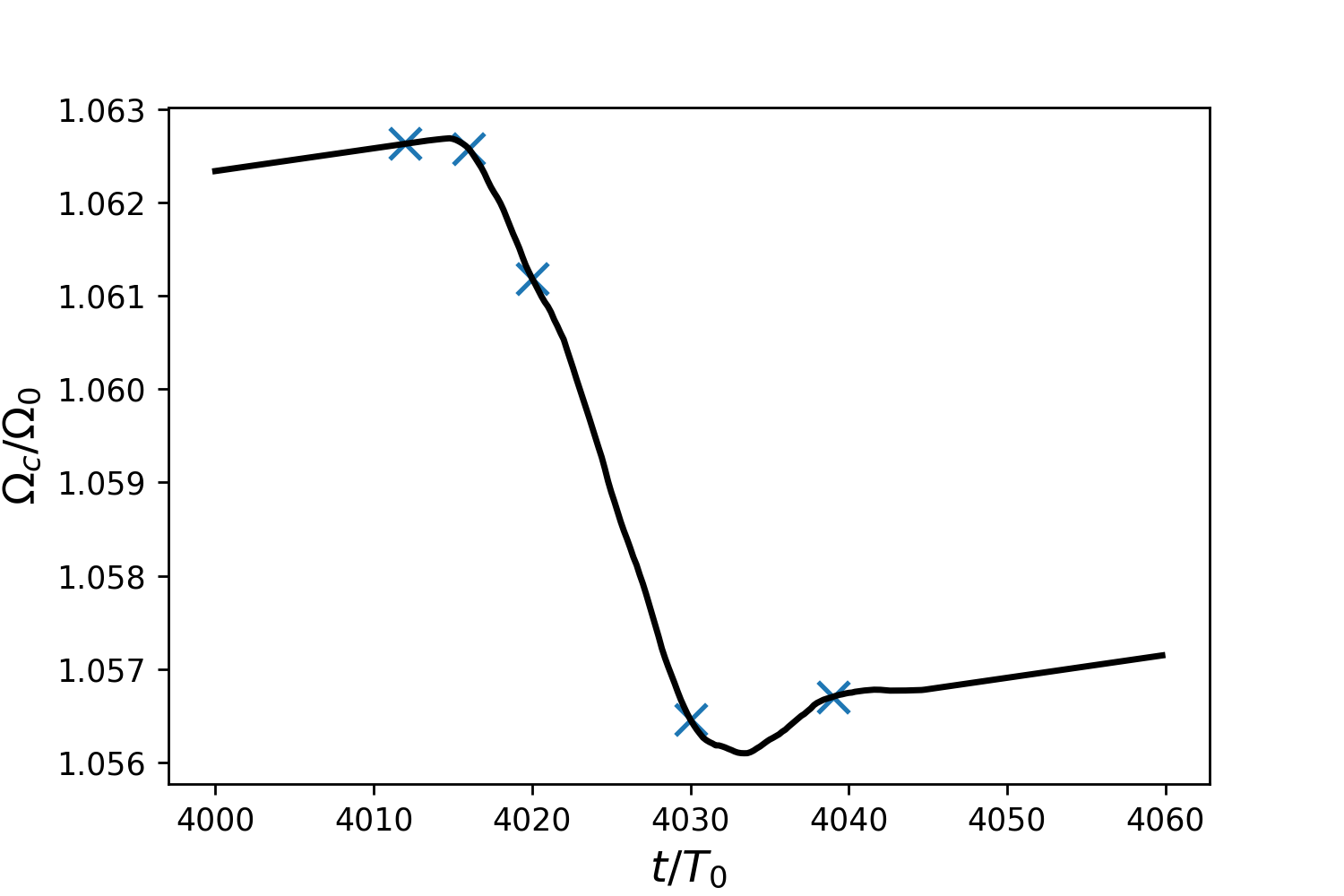}
	\end{center}
	\caption{Vortex motion during an anti-glitch. 
		Top five panels are snapshots from experiment I at  $t/T_0 =$ 4012, 4016, 4020, 4030, 4039 (top to bottom). 
		Grey dots show stationary vortices; red dots show vortices that have moved in the previous time step, black (fading to gray) tracers show the positions of the moving vortices for the 20 previous time steps in the rotating frame of the container.
		Bottom panel: container angular velocity $\OmegaC$ versus time $t$.
		Snapshots from the top five panels are marked with blue crosses.
	}
	\label{fig:avalanche}
\end{figure}

Observing the vortex motion directly, we see that anti-glitches in an accelerating container occur in much the same way as glitches in a decelerating container, but in reverse.
Single vortices unpin and move inwards, causing knock-on unpinning of other vortices which also move inwards.
Figure \ref{fig:avalanche} depicts the motion of vortices during the largest anti-glitch in experiment I, which occurs at $t \approx 4000 T_0$ and has $\Delta \OmegaC / \Omega_0 = -5.9 \times 10^{-3}$, cf. Figures 5 and 9 in 
\citet{Howitt2020}.
%
The top five panels show how the avalanche progresses, with $\OmegaC$ versus $t$ shown in the bottom panel.
The moving vortices are indicated by red dots, with their positions in the previous 20 time steps indicated by black (fading to grey) tracers. 
The stationary vortices are indicated by grey dots.
\begin{itemize}
	\item At $t/T_0 = 4012$, two vortices unpin in the lower right quadrant of the container. 
	Another vortex unpins in the bottom left close to the boundary.
	Neither unpinning event noticeably changes the evolution of $\OmegaC$.
	\item At $t/T_0 = 4016$, more vortices unpin in the lower right quadrant through proximity knock-on and move towards the centre of the container.
	The other unpinned vortex in the bottom left remains unpinned, but does not knock-on other vortices.
	The inward motion of the unpinned vortices changes the sign of $\dot{\Omega}_{\rm C}$.	
	\item At $t/T_0 = 4020$ the avalanche continues in the lower right quadrant. 
	As well as the vortices unpinned by proximity knock-on, some vortices unpin closer to the boundary than the original unpinned vortices, and move inwards towards the depleted region left by the first batch of unpinned vortices.
	The other unpinned vortex in the bottom left region has re-pinned.
	$\dot{\Omega}_{\rm C}$ remains negative. 
	\item At $t/T_0 = 4030$, the unpinned vortices begin to re-pin. 
	As they do so, their trajectory bends clockwise as opposed to the more radial motion earlier in the avalanche.
	$\dot{\Omega}_{\rm C}$ begins to level off.
	\item At $t/T_0 = 4039$ almost all of the vortices that unpinned in the avalanche have re-pinned, and the steady increase of $\OmegaC$ resumes.
\end{itemize}

Figure \ref{fig:avalanche} demonstrates that if vortex avalanches are responsible for glitches in electromagnetically-braking pulsars then the same mechanism can cause anti-glitches in accretion-powered pulsars in principle.

\subsection{Size and waiting time statistics}
\label{subsec:kdes}

While the physical mechanism causing anti-glitches and glitches may be the same, the statistical properties of the glitches and anti-glitches in equivalent simulations differ.
Generally, anti-glitches are smaller in magnitude and more frequent than the glitches.
The smallest anti-glitch in experiment I has 
$\vert \Delta \OmegaC \vert / \OmegaC= 1.1 \times 10^{-6}$ and the largest anti-glitch has 
$\vert \Delta \OmegaC \vert / \OmegaC = 5.9 \times 10^{-3}$, while the 
smallest glitch in experiment II has $\vert \Delta \OmegaC \vert / \OmegaC = 4.7 \times 10^{-6}$ and the largest glitch has $\vert \Delta \OmegaC \vert / \OmegaC = 8.7 \times 10^{-3}$.
There are 176 anti-glitches and 101 glitches.
 
\begin{figure}
	\begin{center}
		\includegraphics[scale=0.5]{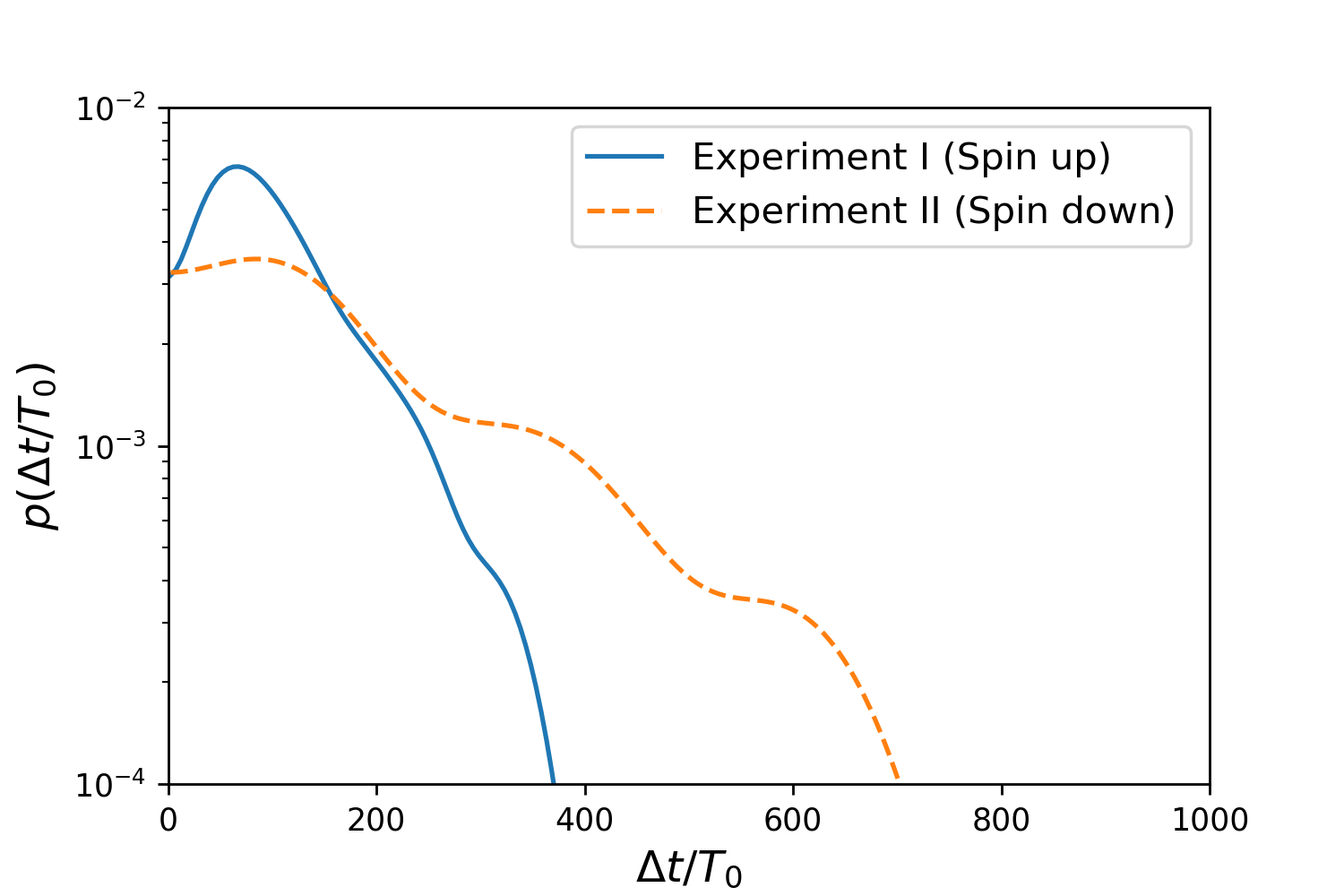}
		\includegraphics[scale=0.5]{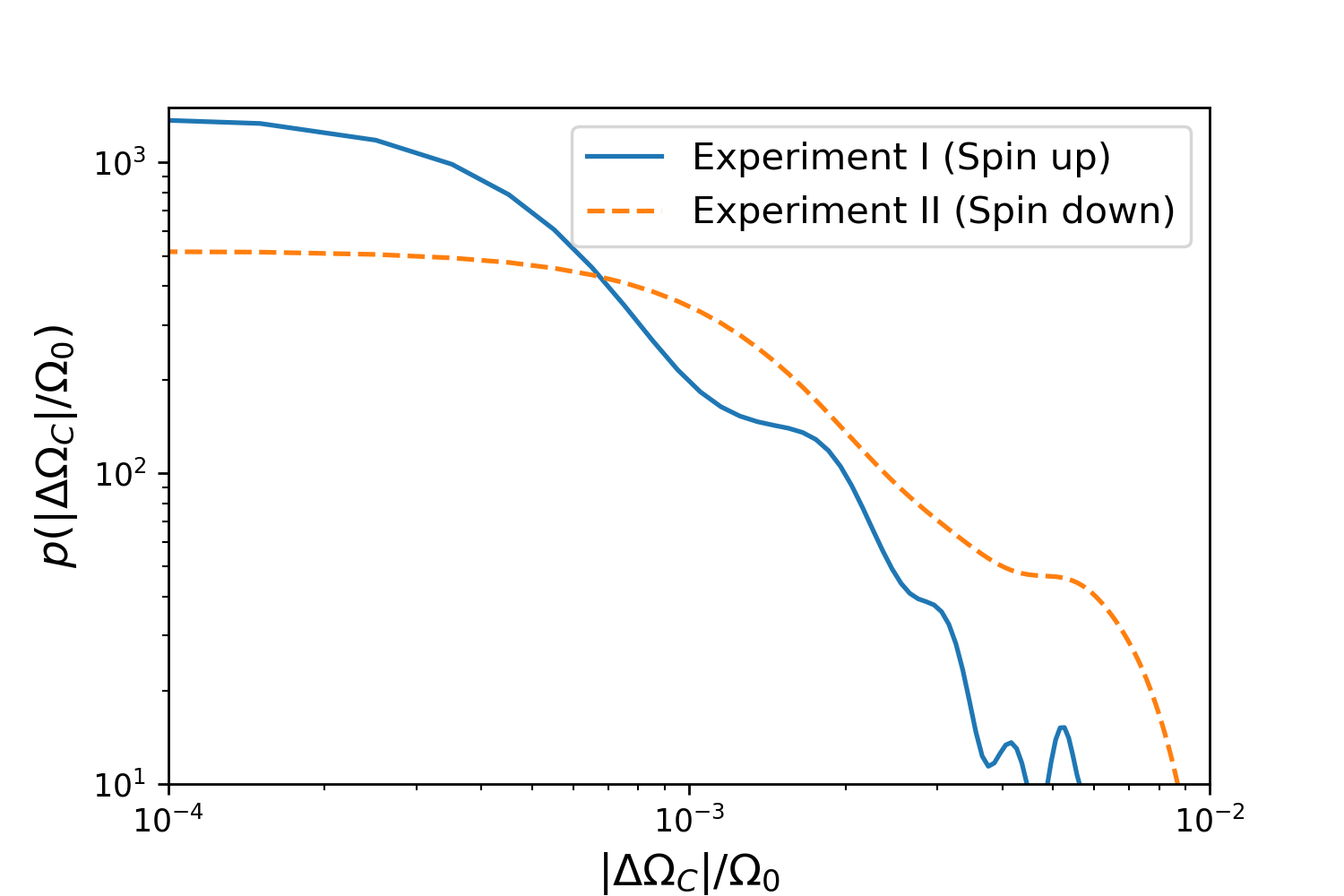}
		\caption{Kernel density estimates of the glitch waiting time (top panel) and size (bottom panel) PDFs for experiment I (blue curve) and experiment II (orange dashed curve). }
		\label{fig:kdes}
	\end{center}
\end{figure}

Figure \ref{fig:kdes} shows kernel density estimates (KDEs) of the glitch/anti-glitch size and waiting time PDFs in experiments I and II.
For sizes, we look at the PDF of the absolute value $\vert \Delta \OmegaC \vert / \OmegaC$ for ease of comparison.
The size and waiting time PDFs for both experiments I and II have similar shapes, however, the distributions for anti-glitches are steeper.
A two-sample Kolmogorov-Smirnov (KS) test returns a $p$-value of $2.8 \times 10^{-4}$  for the waiting times and $4.4 \times 10^{-3}$ for the sizes.
This means that the null hypothesis that the PDFs are the same for each simulation is rejected for both the sizes and waiting times.

\subsection{Why do spin up and spin down differ?}
\label{subsec:differences}

While initially surprising, the result that anti-glitches tend to be smaller and more frequent can be at least partially explained by the interplay between local and global dynamics of the pinned vortex array.
The argument presented in this section is reasonable but not definitive, because the far-from equilibrium dynamics are too complex for an analytic proof. We explore other possible causes of the glitch/anti-glitch size/waiting time discrepancy in Appendix \ref{appendix} below.

The angular momentum of the superfluid is given by 
\citep{Fetter1966,Howitt2020}
\begin{equation}
	L_{\rm S} = k \sum_{i=1}^{N_{\rm v}} (R^2 - r_i^2) \quad ,
\end{equation}
where $k$ is a constant with units of kg s$^{-1}$ and $r_i$ is the radial coordinate of a vortex.
If a vortex moves from 
$r_i$ to $r_i + \Delta r$, 
the change in the angular velocity of the container, by conservation of angular momentum, is given by
\begin{equation}
\Delta \OmegaC \propto \Delta r^2 + 2 r_i \Delta r \quad ,
\label{eq:angular momentum change}
\end{equation}
i.e. $\vert \Delta \OmegaC \vert$ is less for vortices moving inward than outward by the same radial distance.

\begin{figure}
	\centering
	\includegraphics[scale=0.5]{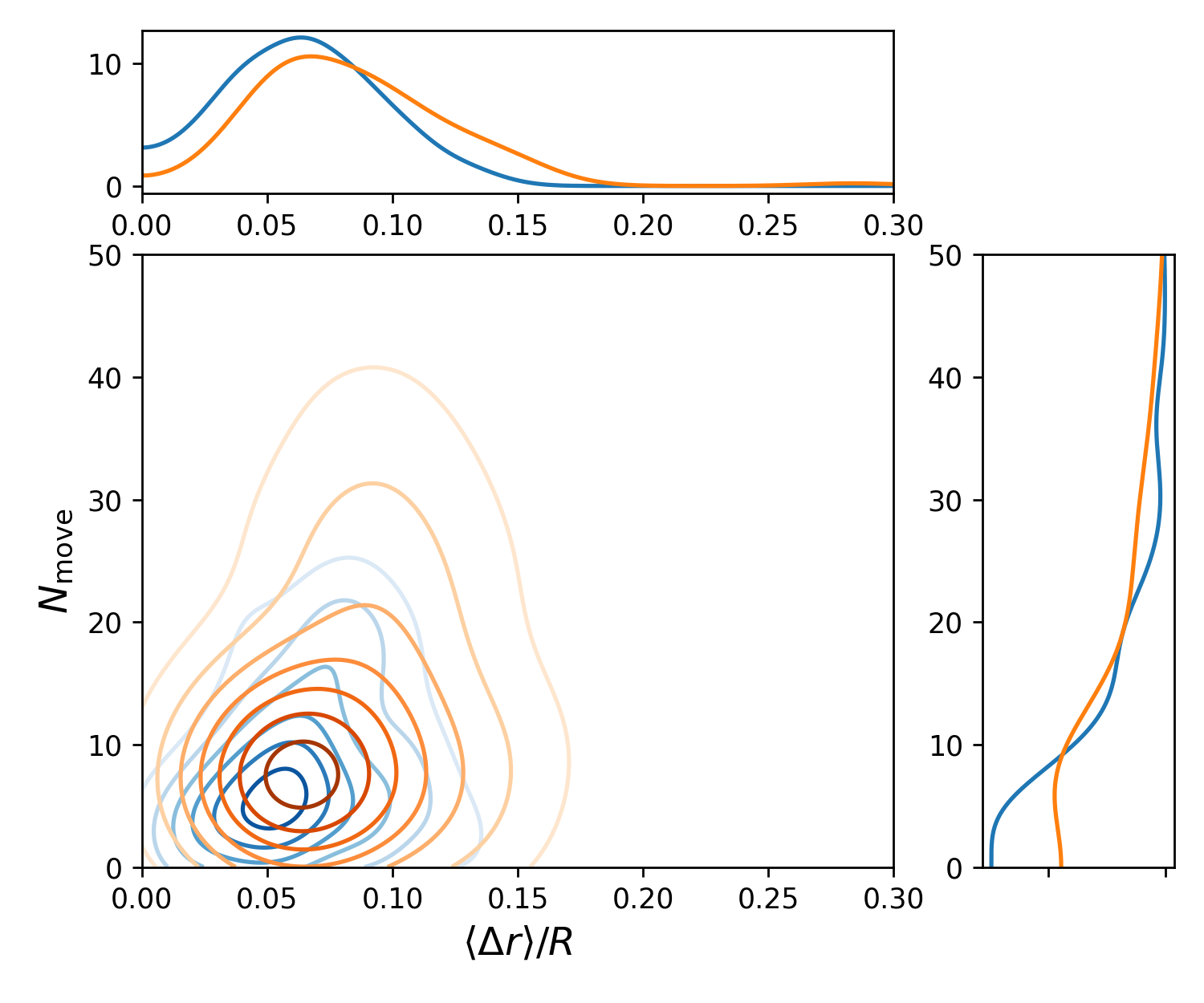}
	\caption{Joint PDF of mean distance travelled by vortices in avalanches $\langle \Delta r \rangle$ and number of vortices in each avalanche $N_{\rm move}$ for experiment I (blue contours) and experiment II (orange contours).
	1D projections of the individual PDFs are shown in the margins.}
\label{fig:mean vortex distance}
\end{figure}

In Figure \ref{fig:mean vortex distance},
we show the joint PDF of the mean distance travelled by vortices in each avalanche, 
$\langle \vert \Delta r \vert \rangle$,
in experiments I and II, as well as the number of vortices that move a distance $\Delta r > a$, denoted by $N_{\rm move}$.
Figure \ref{fig:mean vortex distance} shows that the PDFs of 
$\langle \vert \Delta r \vert \rangle$ and $N_{\rm move}$
are similar in experiments I and II. 
Avalanches involve the same number of vortices moving the same radial distance before re-pinning, on average.
The reduced angular momentum transfer from inward-moving avalanches is therefore primarily responsible for the reduced size of anti-glitches.

This is only half the story, however. 
One might wonder why the inward-moving vortices in experiment I do not travel further in order to transfer as much angular momentum as in experiment II.
When a vortex unpins, and where it subsequently re-pins, are determined by the local stress at the location of the vortex
\citep{Haskell2016}. 
In hydrodynamical descriptions of superfluid dynamics
[e.g.\citet{Hall1956,Barenghi1983}],
this stress $S$ is proportional to the velocity lag between a viscous `normal' fluid component with velocity $\mathbf{v}_n$ and an inviscid superfluid component with velocity $\mathbf{v}_s$, i.e.
$S \propto \vert \mathbf{v}_s - \mathbf{v}_n \vert$.
Let us consider a single vortex at radial coordinate $r$ which moves to $r + \Delta r$. 
Ignoring the local contributions to the superfluid velocity from the vortices in its immediate neighbourhood (an appropriate approximation given that we are trying to explain only the average behaviour of many vortices across many avalanches), we can take both the superfluid and normal fluid velocity fields to be azimuthal.
The normal fluid co-rotates with the container, so one has
$v_n = \OmegaC r$.
The superfluid velocity, by the Feynman condition is
$v_s = \kappa N(r) / r$, where $N(r) \propto r^2$ is the number of vortices within a circle of radius $r$ centred on the origin. 
Thus the change in the stress for a vortex moving from $r$ to $r+ \Delta r$ is
\begin{equation}
\ S = S(r + \Delta r) - S(r) 
\propto  \Delta r
\label{eq:stress change}
\end{equation}
The symmetry of Equation \eqref{eq:stress change} with respect to $\Delta r$ explains why vortices move the same distance on average in avalanches in experiments I and II, while the asymmetry in Equation \eqref{eq:angular momentum change} explains why anti-glitches are smaller in magnitude than glitches.

The fact that anti-glitches are smaller than glitches does not mean that the differential rotation is greater in accretion-powered pulsars. 
In both experiments I and II, after the onset of avalanches the magnitude of the spin-up/down rate (averaged over the entire simulation) reduces to approximately half its initial value, as shown in Figure \ref{fig:spinup_frequency}, reflecting the fact that avalanches couple the superfluid moment of inertia ($I_{\rm rel}=1$) to that of the crust.
While anti-glitches are smaller than glitches, the corresponding reduced waiting times mean that they transfer the same amount of angular momentum over a long-enough time scale.

It is unlikely that the statistical asymmetry between glitches and anti-glitches is observable in pulsars.
Superfluid vortices likely \textbf{pin} only in the inner crust/outer core region at $r \gtrsim 0.9R$, and the typical glitch size of $\Delta \OmegaC / \OmegaC \sim 10^{-6}$ implies $\Delta r \ll r_i$, so the second term in Equation \eqref{eq:angular momentum change} dominates. 
It may however be observable in terrestrial experiments with superfluids or Bose-Einstein condensates, which necessarily involve fewer vortices than the $\sim 10^{15}$ in pulsars and stronger torques.

As well as the simulations with $N_{\rm v} = 2000$ presented here, we also performed a suite of comparable simulations on systems with $500 \leq N_{\rm v} \leq 5000$. 
The results from these simulations are qualitatively similar to those discussed in this section and are not displayed for brevity.

\subsection{Model variations}
\label{subsec:model variations}

As in 
\citet{Howitt2020},
we run a suite of spin-up simulations varying several of the input parameters, namely the pinning strength $V_0$, the pinning site separation $a$, the dissipation angle $\phi$ and the external spin up rate $N_{\rm ext}$.
In each case, we keep all the other parameters at their same value as in the default simulation.
As well as the default values given at the beginning of Section \ref{sec:results}, we run simulations with $V_0=1000$ and $V_0 = 4000$; $a=0.025R$ and $a=0.005R$, corresponding to pinning site-vortex abundance ratios of $\approx 1$ and $\approx 100$ respectively; $\phi=0.01$ rad and $\phi = 0.5$ rad; and $N_{\rm ext} = -5 \times 10^{-4} \Omega_0 / T_0$ and $N_{\rm ext} = - 2 \times 10^{-3} \Omega_0 / T_0$.

In Table \ref{tab:model variations} we show the qualitative effect of adjusting each of the parameters on the mean absolute glitch size 
$\langle \vert \Delta \OmegaC \vert \rangle$ and mean waiting time 
$\langle \Delta t \rangle$ in both the spin-up simulations in this paper and (by way of comparison) the spin-down simulations from 
\citet{Howitt2020}.
\begin{table*}
	\centering
	\begin{tabular}{|c|c|c|c|c|}
		\hline 
		Parameter & $\langle \Delta t/T_0 \rangle$ (spin up) & $\langle \Delta t/T_0 \rangle$ (spin down) &  $\langle \vert \Delta \OmegaC \vert \rangle$ (spin up) &  $\langle \vert \Delta \OmegaC \vert \rangle$ (spin down) \\ 
		\hline 
		$V_0$ & $+$ & $+$ &  $+$ & $+$  \\ 
		$a$ & . & $+$ & . & .   \\ 
		$\phi$ & . & . &  . & .  \\ 
		$\vert N_{\rm ext} \vert$ & $-$ & $-$ & . & .  \\ 
		\hline 
	\end{tabular} 
\caption{Effect of varying simulation parameters (first column) on vortex avalanche statistics: mean waiting time $\langle \Delta t/T_0 \rangle$ between anti-glitches (second column) and glitches (third column); and mean size $\langle \vert \Delta \OmegaC \vert \rangle$ of anti-glitches (fourth column) and glitches (fifth column).
Symbols $+$ ($-$) indicate that $\langle \Delta t \rangle$ or $\langle \vert \Delta \OmegaC \vert \rangle$ increases (decreases) relative to the default value in section \ref{subsec:avalance dynamics} as the parameter in column one increases;
``." indicates no consistent effect.
}
\label{tab:model variations}
\end{table*}

Table \ref{tab:model variations} shows that the anti-glitches have a similar response to parameter variations as glitches.
As expected, our findings are largely the same as in 
\citet{Howitt2020}.
Increasing the pinning strength $V_0$ makes vortices harder to unpin, so fewer avalanches occur.
When vortices unpin they are more stressed, leading to more knock-on and hence larger glitch sizes.
Increasing the pinning site separation $a$ (i.e. decreasing the pinning site density) increases the waiting time in the spin down case without affecting the glitch size, while no effect is observed in spin up simulations.
The nominal difference between the spin-up and spin-down simulations versus $a$ is marginal and may be related to the inability of the glitch-finding algorithm to resolve small avalanches involving few vortices.
Changing the dissipation angle $\phi$ from the default value $\phi = 0.1$ rad in either direction results in more, smaller glitches and anti-glitches. 
The cause is complicated and discussed in detail in Section 5.4 of 
\citet{Howitt2020}.
Increasing/decreasing the external torque $\vert N_{\rm ext} \vert$ increases/decreases the number of avalanches because vortices accumulate stress faster/slower, but the avalanche sizes are unaffected.

We also study forward and backward cross-correlations between sizes and waiting times for all of our model variations, cf. Section 5.5 in
\citet{Howitt2020}. 
For the default parameters (experiment I), the Spearman rank coefficient for the forward correlation, i.e. the correlation between size and subsequent waiting time is  $\rho_+ = 0.54$ compared to $\rho_+=0.44$ for experiment II. 
The backward correlation, i.e. the correlation between size and the preceding waiting time, is $\rho_- = -0.10$ for experiment I compared to $\rho_-=0.03$ for experiment II. 
We refrain from including correlation coefficients from all simulations here, as the results are basically the same as those shown in 
\citet{Howitt2020}.
We see weak but statistically significant forward correlations for all parameter variations except weak dissipation, and we see no statistically significant backwards correlations.
This stands in contrast to the measured correlations in glitching pulsars, where few significant forward correlations are observed
\citep{Melatos2018}.

\section{Idealizations in the point vortex model}
\label{sec:idealizations}

The point vortex model described in Section \ref{sec:method} makes a number of approximations and simplifications.
It is not intended to be a fully realistic model of a neutron star, given the many physical and astrophysical uncertainties in the problem. 
Several of the idealizations involved are discussed thoroughly by 
\citet{Haskell2015}
and in Section 7 of 
\citet{Drummond2018}
in the Gross-Pitaevskii context. 
In this section, we discuss three issues of special relevance to the point vortex model in this paper: the cylindrical geometry, extrapolation to higher $N_{\rm v}$, and the physics of vortex pinning in the presence of magnetic flux tubes. 

\subsection{Cylindrical geometry}
\label{subsec: cylindrical geometry}

It is natural to ask whether the artificial cylindrical geometry in this paper behaves differently to the spherical geometry expected in a neutron star, at least as far as anti-glitch observables are concerned.
There are three key issues.
\begin{enumerate}[(a)]
	\item The motion of a neutron star vortex is insensitive to conditions at its endpoints (including the shape of the boundary), because vortices are long and thin. The length of a vortex ($\sim 10\, {\rm km}$) is much greater than its diameter ($\sim 10\,{\rm fm}$), if the simple Feynman-Onsager picture of a rotating superfluid is applied uncritically to a neutron star 
	\citep{Haskell2015},
	without allowing for the macroscopic and microscopic imperfections which are likely to exist, e.g.\ grain boundaries and lattice dislocations in the crust, compositional variations, gravitational stratification, and inhomogeneous superfluid phases. 
	In the simple picture, therefore, the middle of a $\sim 10$ km vortex (near the star's equatorial midplane) is not affected meaningfully by what happens at the boundaries, where vortices terminate.
	To an excellent approximation, every vortex terminates at a wall which is perpendicular to its length, i.e. normal at the point of contact, because the radius of curvature of the star ($\sim 10$ km) is much greater than the vortex diameter ($\sim 10$ fm) and even the vortex separation ($\sim 1$ mm).
	The physical decoupling between most of a vortex and its endpoints is even more pronounced, when the realistic inhomogeneities listed above are considered. 
	One way to visualize this is that Kelvin waves propagating along a vortex from the endpoints encounter many complicated scattering centers and refractive index variations, which cause them to lose ``memory'' of the conditions at the boundary. 
	In this sense, point vortices are a better approximation than $\sim 10\,{\rm km}$ vortices. 
	Their effective (as opposed to nominal) length, which is much less than $10\,{\rm km}$, is discussed in point (b) below.
	\item  Superfluid vortices are not perfectly rigid. 
	Their rigidity is quantified by a stiffness parameter 
	$T_{\rm v} r_0 / (F_{\rm pin} a)$, where $T_{\rm v}$ is the vortex tension 
	\citep{Hall1956,Barenghi1983}, $r_0$ is the pinning site radius (roughly equivalent to $\xi$ in our notation), $F_{\rm pin}$ is the characteristic pinning force, and $a$ is the pinning site separation.
	The stiffness is dimensionless and gives roughly the number of pinning sites across which the vortex moves coherently as a rigid bar 
	\citep{Link1991,Link1993,Seveso2016,Tsubota2017}.
	If $T_{\rm v}$ is relatively low, with $T_{\rm v} r_0 \lesssim F_{\rm pin} a$, vortex unpinning occurs through single-site breakaway; if $T_{\rm v}$ is relatively high, with $T_{\rm v} r_0 \gtrsim F_{\rm pin} a$, vortex unpinning occurs through multi-site breakaway
	\citep{Link1991}.
	Either way we have $T_{\rm v} r_0 / (F_{\rm pin} a) \lesssim 10^2$ according to Equation (3.9) in 
	\citet{Link1993}, so even the stiffest vortex is rigid on microscopic scales only. 
	Therefore a 10-km long vortex behaves like a ``gas" of $\gtrsim 10^{15}$ independent point vortices, each moving in a plane, as in our simulations, and only a tiny fraction interact with the walls. 
	This is even truer in the realistic case of a polarized vortex tangle, which we do not treat in this paper
	\citep{Tsubota2003,Peralta2006b,Andersson2007}.  
	\item Experiments by 
	\citet{Tsakadze1980}
	involving pinned vortices in decelerating containers show spasmodic, glitch-like dynamics for both cylindrical and spherical containers, without much qualitative difference. 
	This is consistent with the hydrodynamic theory in 
	\citet{vanEysden2010,vanEysden2013}
	where spherical and cylindrical containers are compared. 
	Geometry does exert a subtle influence on post-glitch recovery time-scales (of order weeks to months), because Ekman circulation occurs differently in cylinders and spheres, but the differences are small and irrelevant to the fast vortex unpinning discussed in the present paper, which occurs over $\lesssim |\Omega_{\rm S} - \Omega_{\rm C}|^{-1} \sim 1\, {\rm s}$ 
	
\end{enumerate}

\subsection{Extrapolation to higher  $N_{\rm v}$}
\label{subsec:extrapolation}

The simulations in this paper track $N_{\rm v} \lesssim 10^4$ vortices, compared to $N_{\rm v} \gtrsim 10^{15}$ vortices in a realistic, accreting neutron star. 
While this is unavoidable due to computational limitations, it is reasonable to ask to what extent the results of the paper extrapolate to realistic $N_{\rm v}$. 
More work is needed to answer this question definitively, but for now the key criterion seems to be the number of pinning sites per vortex: as long as there are many pinning sites per vortex, extrapolation is relatively safe. 
Intuitively, if there are many pinning sites in its vicinity, a vortex pins randomly and without preference to any one of them, at a location approximately where it would stand in a regular Abrikosov array, and this holds true whether there are $10^2$ (in a simulation, say) or $10^{18}$ (in reality) pinning sites in its vicinity. 
Evidence collected to date from quantum mechanical Gross-Pitaevskii simulations 
\citep{Warszawski2011, Melatos2015}
and hydrodynamic $N$-body simulations 
\citep{Howitt2020} supports this conclusion, as does the analogy with self-organized critical systems 
\citep{Jensen1998};
vortex avalanches exhibit the same properties qualitatively, as long as there are more pinning sites than vortices. 
Encouragingly this is true for a range of pinning sites per vortex (approximately 1 to $10^2$), even though the $N$-body simulations do not include the acoustic knock-on process observed in Gross-Pitaevskii simulations 
\citep{Warszawski2012a}.
We are currently investigating analytic approaches to predict glitch scalings versus $N_{\rm v}$ for $N_{\rm v} \gg 1$, but this is a hard problem in far-from-equilibrium statistical mechanics and remains unsolved
\citep{Jensen1998}. 
The number of pinning sites per vortex greatly exceeds unity in a neutron star, whether the pinning occurs at nuclear lattice sites or magnetic flux tubes (see Section \ref{subsec:flux tube pinning}). 
The dimensional value of quantities like $\langle \Delta t \rangle$ in a neutron star depends on the dimensional value of system-specific quantities like $N_{\rm ext}$, of course.
	
\subsection{Magnetic flux tube pinning}
\label{subsec:flux tube pinning}

The calculations in this paper are largely agnostic about the pinning microphysics, partly because the avalanche dynamics are insensitive to the microphysics [a general property of other self-organized critical systems as well, e.g. sand piles
\citep{Jensen1998}]
and partly because the microphysics is poorly known 
\citep{Haskell2015}. 
In this paper, the values of parameters like $V_0$, $a$, and $\phi$ are chosen to broadly reflect the situation expected for pinning at nuclear lattice sites, although their dimensionless values are unrealistic due to numerical limitations; see also Sections \ref{subsec: cylindrical geometry} and \ref{subsec: cylindrical geometry}. 
In this section, for the sake of completeness, we explore briefly how the situation changes for pinning at magnetic flux tubes 
\citep{Ruderman1998,Sourie2020}. 
However, we caution that the properties (e.g. number, geometry, and location) of magnetic flux tubes are even less certain in accreting neutron stars than in isolated objects. 
The global dipole moment is typically $\lesssim 10^4$ times weaker in accreting systems, but the local internal fields can be strong and tangled
\citep{Melatos2004}, 
and they may or may not be excluded from certain regions (e.g. the outer core) 
\citep{Vigano2013, Turlione2015}.

The pinning microphysics influences vortex avalanches in two main ways, through the pinning site separation $a$ and dissipation angle $\phi$.
With respect to pinning site separation, in the context of magnetic flux tube pinning, one finds $a = 4 \times 10^{-10} (B/10^{12} \, {\rm G})^{-1/2} \, {\rm cm}$, where $B$ is the surface magnetic field strength.
This is typically greater than the nuclear lattice site separation ($\sim {\rm fm}$) and flux tube core radius $\approx 5 \times 10^{-13} (n_{\rm p}/10^{36} \, {\rm cm^{-3}} )^{-1/2} \, {\rm cm}$, where $n_{\rm p}$ is the proton number density, but it is smaller than the vortex separation. 
Hence flux tube pinning resembles the situation simulated in Section \ref{sec:results} and discussed in Section \ref{subsec:extrapolation}, where pinning sites are more abundant than vortices.
Although the ratio of pinning sites to vortices satisfies $\lesssim 10^2$ out of computational necessity, Gross-Pitaevskii simulations suggest that the event statistics (e.g. size and waiting-time PDFs) do not change when the ratio increases further
\citep{Warszawski2011},
consistent with other self-organized critical systems 
\citep{Jensen1998}.
In two-dimensional simulations, or in idealized three-dimensional systems where the flux tubes and vortices are rectilinear and parallel, the number of vortex-flux tube intersections, $N_{\rm p}$, termed “pinned fluxoids” by 
\citet{Sourie2020}, 
equals the number of vortices. 
However, if the vortices and flux tubes are inclined or tangled, or vortex-fluxoid clusters form 
\citep{Sedrakian1995b},
the number of pinned fluxoids can reach as high as $\sim 10^{13} (B / 10^{12} \, {\rm G}) (\Omega / 10^{3} \, {\rm rad} \, {\rm s}^{-1})^{-1}$ 
\citep{Sourie2020}. 
Near the upper end of this range, $a$ approaches the nuclear lattice site separation.

With respect to the dissipation angle, one has $\tan\phi = {\cal R}$, where ${\cal R}$ is the ratio of the drag force to the Magnus force and can be related to the mutual friction coefficient ${\cal B}\approx {\cal R}$.
\citet{Sourie2020} calculated $10^{-13} \lesssim {\cal B} \lesssim 10^{-3}$ for $1\leq N_{\rm p} \leq 10^{12}$ in the vortex-cluster model 
\citep{Sedrakian1995b}.
The full range implies $\phi \ll 1$, which is the regime simulated in Section \ref{sec:results}. 
Table \ref{tab:model variations} suggests that the event statistics are insensitive to the exact value of $\phi$, as long as it is small, but in fairness we are restricted to $0.01 \leq \phi \leq 0.1$ out of computational necessity and do not probe the lower extremes of the range, viz. $\phi \sim 10^{-13}$, where different behavior is conceivable.

It is worth noting that the avalanche dynamics are insensitive to the specific, microphysical value of $V_0$. 
Self-organized systems tend towards marginal stability: a sand pile fluctuates around a critical slope 
\citep{Jensen1998}, 
and likewise the crust-superfluid angular velocity lag in a neutron star fluctuates around a critical value set by the balance of pinning, drag, and Magnus forces. 
$V_0$ influences the critical state, i.e. the baseline (and hence quantities like the mean waiting time), but it does not play much role in determining the shapes of the size and waiting-time PDFs, for example. 
Even though $V_0$ differs in general for pinning at nuclear lattice sites and magnetic flux tubes, one expects the glitch statistics to be similar, as in other self-organized critical systems 
\citep{Jensen1998}.
One subtle issue, however, is whether the point-like pinning potential in this paper is suitable for flux-tube pinning, which involves a longer-range current-current component as well as a short-range density-density component 
\citep{Srinivasan1990}. 
It is also unclear whether vortices and flux tubes are tangled or rectilinear when they interact locally; the relevant mesoscopic length-scales cannot be resolved for realistic neutron star parameters by Gross-Pitaevskii and Ginzburg-Landau simulations at present 
\citep{Drummond2017, Drummond2018}.

\section{Conclusion}
\label{sec:discussion}

We perform simulations of 500---5000 pinned superfluid vortices in accelerating containers.
We find that vortex avalanches interrupt the secular increase of the angular velocity to produce anti-glitches for a wide range of physical parameters. 
The results match those demonstrating glitches in decelerating containers from
\citet{Howitt2020}.
The size and waiting time statistics of anti-glitches are qualitatively similar to glitches in decelerating containers. 

Interestingly, anti-glitches in our simulations are smaller and more frequent than glitches, an effect we believe is due to reduced angular momentum transfer between the superfluid and container when vortices move inward rather than outward. 
Such an effect is unlikely to be observable in neutron stars, where\textbf{ one has} $\Delta r \ll r_i$ [cf. Equation \eqref{eq:angular momentum change}], but may be tested in terrestrial experiments on quantized vortices in superfluids and Bose-Einstein condensates.

The anti-glitches in NGC 300 ULX-1 suggest a direct analogue with standard glitching pulsars, and hence offer an opportunity to test the applicability of the vortex avalanche model in a new regime.
The results in Section \ref{sec:results} and Figure \ref{fig:avalanche} confirm that, from a microphysical standpoint, inward-propagating vortex avalanches are a viable explanation for anti-glitches in accelerating neutron stars, at least in principle
\citep{Ray2019}.
Further timing campaigns of accelerating ULX pulsars will hopefully discover more anti-glitches in the future, enabling closer comparison to the population of glitches in isolated neutron stars.
It is worth noting, however, that these objects are often only observable in outburst, making construction of phase-coherent timing solutions challenging.
Another point worth noting is that not all accreting pulsars are accelerating
\citep{Bildsten1997}.
This is because the hydromagnetic accretion torque can take either sign depending on the fastness parameter
\citep{Ghosh1977}.

In this paper, we do not seek to model anti-glitches in magnetars 
\citep{Dib2008}. 
The vortex equations of motion \eqref{eq:total velocity}--\eqref{eq:y velocity} do not include hydromagnetic forces, which are likely to be important dynamically in magnetars. 
For example, the kinetic energy per unit volume associated with the crust-superfluid lag, $\rho |{\bf v}_{\rm s} - {\bf v}_n|^2 \sim 10^{23} \, {\rm erg\,cm^{-3}}$, and the comparable pinning energy per unit volume, are much smaller than the magnetic energy per unit volume, $\sim 10^{29}\,{\rm erg\,cm^{-3}}$ 
\citep{Glampedakis2011,Mereghetti2015}. 
However, some of the phenomenology of magnetar anti-glitches resembles loosely the results in Section \ref{sec:results}, so it is worth generalizing the treatment in this paper to include magnetic fields in future work, e.g. pinning to superconducting magnetic flux tubes 
\citep{Sidery2009, Drummond2017,Drummond2018}.
Intriguing examples of magnetar anti-glitch phenomenology include the following.
\citet{Archibald2013} reported an anti-glitch in the magnetar 1E 2259+586.
The anti-glitch was detected as a discontinuity in the timing model between two observations taken with the Swift X-ray telescope at 14 April 2012 and 28 April 2012.
On 21 April 2012, the Fermi Gamma-ray Burst Monitor detected a 36ms hard X-ray burst from the same object, possibly coincident with the anti-glitch.
Swift monitoring subsequent to 28 April 2012 measured an increase in the $2-10$ keV X-ray flux of a factor of $\approx 2$, which then decayed following a power law in time.
These associated radiative changes are similar to other glitches in magnetars
\citep{Dib2008,Dib2014}, including one other spin up glitch in 1E 2259+586
\citep{Kaspi2003}, 
and unlike radio pulsar glitches.
Following the anti-glitch, the spin-down rate of the magnetar approximately doubled for $\sim 100$ d before returning to its pre-glitch value following a second glitch.
The sign of the second glitch is uncertain, 
\citet{Archibald2013}
considered two models, one with an anti-glitch followed by a glitch, the other with an anti-glitch followed by a second anti-glitch, and found no statistical preference for either, but preferred both to a model with only a single anti-glitch.
More recently, a spin-up and spin-down glitch have been reported in the same object
\citep{Younes2020},
both without any associated pulse profile changes or enhanced X-ray flux.
As 1E 2259+586 experiences glitch-like timing irregularities of both signs, and is spinning down, it is challenging to reconcile anti-glitches in this object with the vortex avalanche model, and it is possible that they are instead due to changes in the internal magnetic field
\citep{Garcia2015,Mastrano2015}.

\section*{Acknowledgements}

GH and AM acknowledge support from the Australian Research Council (ARC) through the Centre of Excellence for Gravitational Wave Discovery (OzGrav) (grant number CE170100004) and an ARC Discovery Project (grant number DP170103625)

\section*{Data Availability}

Simulation data and code used in this paper can be made available upon request by emailing the corresponding author.



\bibliographystyle{mnras}
\bibliography{antiglitches}




\appendix
\section{Statistical Discrepancy between glitches and anti-glitches}
\label{appendix}

Here we explore and discard some plausible alternative explanations for the discrepancy in event statistics between glitches and anti-glitches discussed in Section \ref{subsec:differences}. 
In Section \ref{subsec:time reversal} we look at whether it is due to the differing values of $N_{\rm v}$ in experiments I and II.
In Section \ref{subsec:avalanche location}, we look at where in the container avalanches originate.
In Section \ref{subsec:spatial correlations} we test for spatial correlations in the distribution of pinned vortices.
In Section \ref{subsec:vortex creep} we look at whether it is due to perturbations to the vortex array caused by the addition of vortices in spin-up simulations.
None of our investigations reveal an obvious cause.

\subsection{Time-reversed simulations}
\label{subsec:time reversal}

\begin{figure}
	\begin{center}
		\includegraphics[scale=0.5]{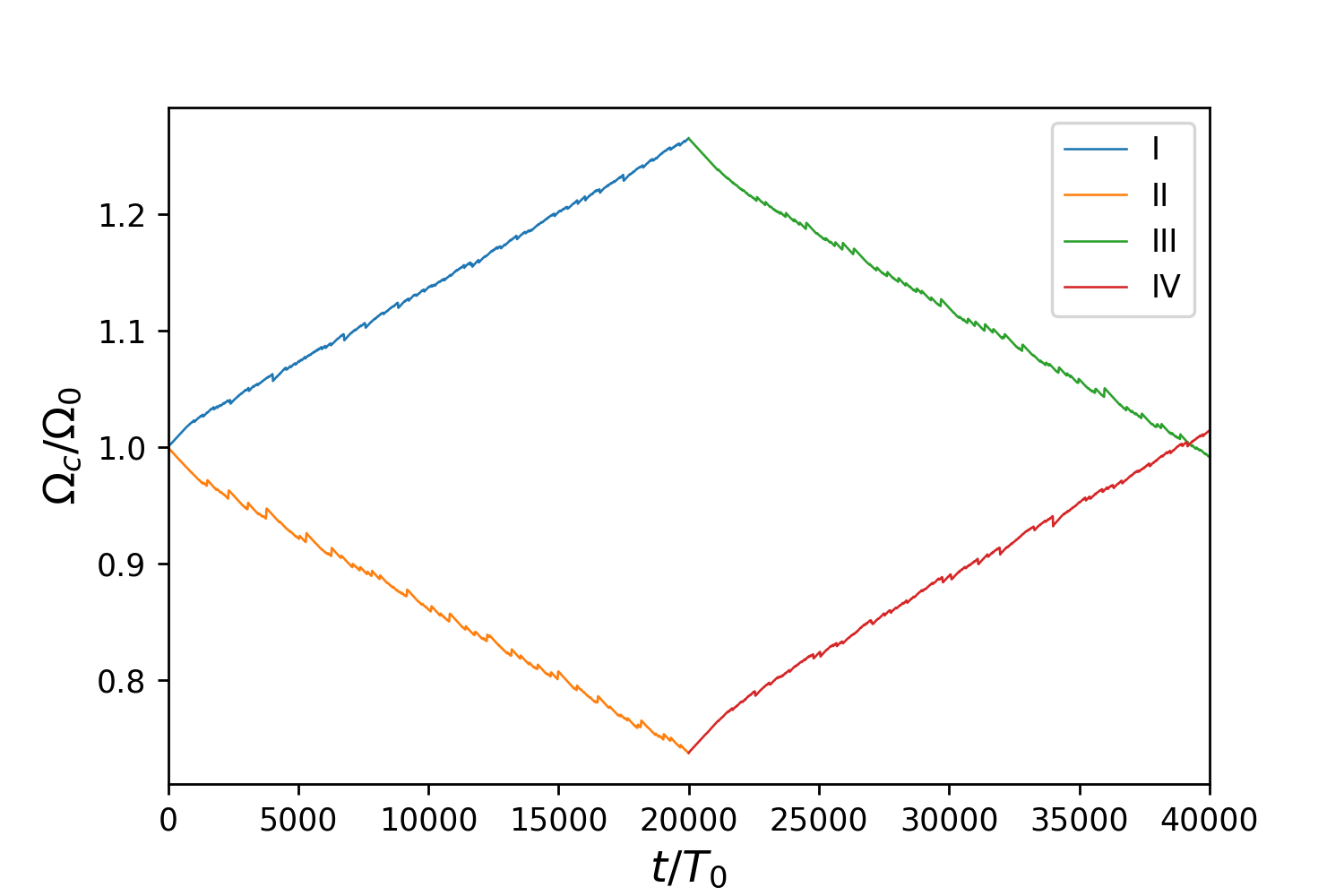}
		\caption{Angular velocity $\OmegaC$ versus time for experiments I (blue curve), II (orange curve), III (green curve) and IV (red curve).}
		\label{fig:all omegas}
	\end{center}
\end{figure}

Both experiments I and II begin with the same initial vortex configuration, but in experiment I $N_{\rm v}$ increases from 1943 at $t=0$ to 2361 at $t/T_0 = 2 \times 10^4$, while in experiment II $N_{\rm v}$ decreases to 1498 by $t/T_0=2 \times 10^4$.
In general, we do not expect a difference in $N_{\rm v}$ of this magnitude to affect the statistics of the sizes and waiting times.
However, to investigate the possibility that this may be the source of the differences in the PDFs between the two simulations, we run time-reversed simulations of each experiment. 
In experiment III, we use the state of the vortex array from experiment I at $t=2 \times 10^4 T_0$ as the initial condition and run for $ 2 \times 10^4 \leq 4 \times 10^4$ with $N_{\rm ext} = -10^{-3} \Omega_0 / T_0$.
In experiment IV, we use the state of the vortex array from experiment II at $t=2 \times 10^4 T_0$ as the initial condition and run for $ 2 \times 10^4 \leq 4 \times 10^4$ with $N_{\rm ext} = 10^{-3} \Omega_0 / T_0$.
The initial and final values of $N_{\rm v}$ and $\OmegaC$ are shown in Table \ref{tab:numerical experiments}.
The evolution of the angular velocity $\OmegaC$ in all experiments is shown in Figure \ref{fig:all omegas}.
The \textbf{latter} figure shows that after changing the sign of the torque, the system undergoes a short period of steady spin down before avalanches begin again, now with the opposite sign as in the initial simulation.
Experiment III has 116 glitches, and experiment IV has 166 anti-glitches. 

\begin{figure}
	\begin{center}
		\includegraphics[scale=0.5]{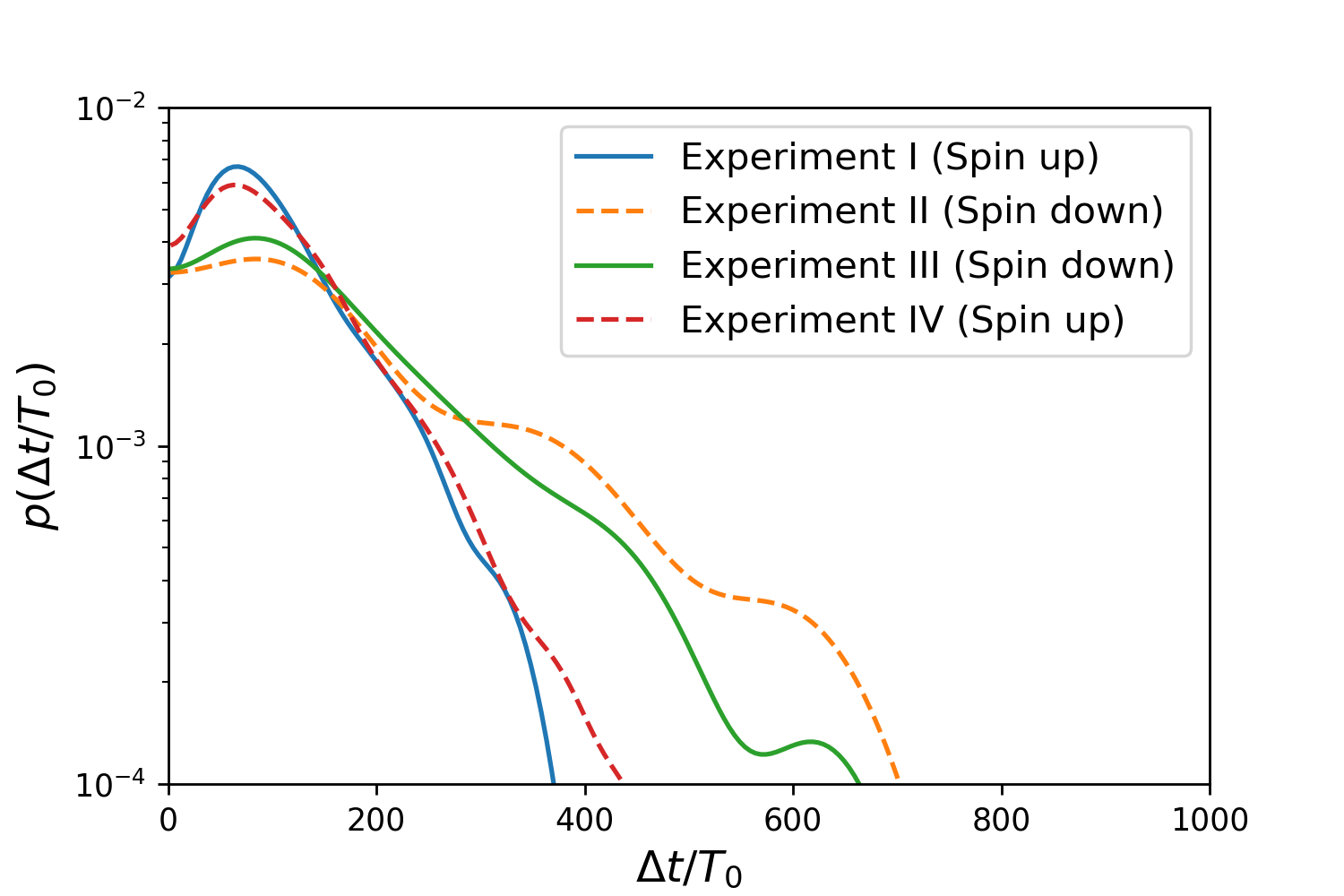}
		\includegraphics[scale=0.5]{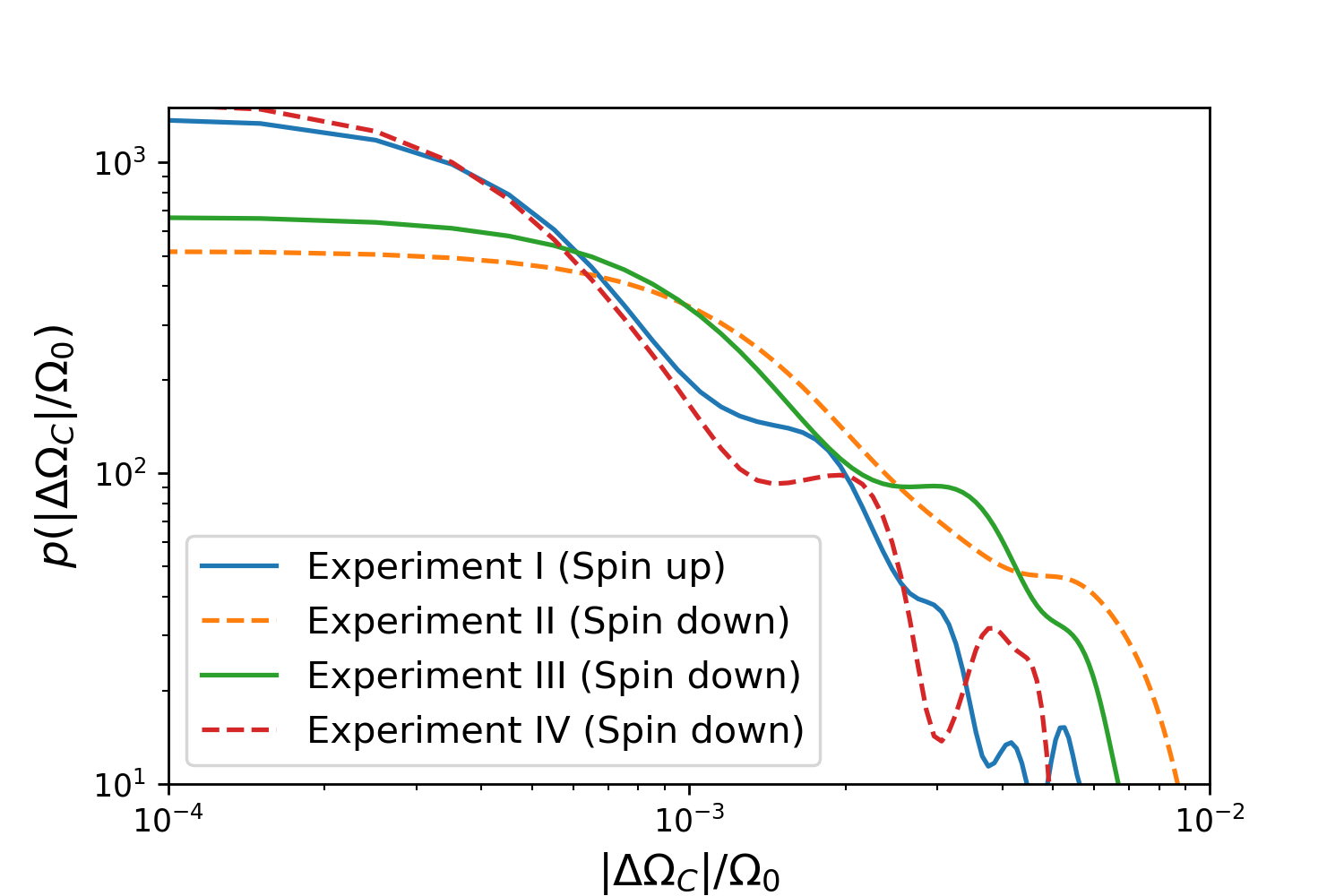}
		\caption{KDEs of the glitch size and waiting time PDFs for experiments I (blue curve), II (orange dashed curve), III (green curve) and IV (red dashed curve).}
		\label{fig:kdes reversed}
	\end{center}
\end{figure}

If the discrepancy in size and waiting time PDFs seen in Figure \ref{fig:kdes} is due to the difference in $N_{\rm v}$, then we expect that the size and waiting time PDFs of the torque-reversed spin up/down simulations should be closer to the spin up/down simulations they extend rather than the initial simulations with the same torque.
The KDEs for the size and waiting time PDFs for all four experiments are shown in Figure \ref{fig:kdes reversed}.
Figure \ref{fig:kdes reversed} shows that the simulations with the same sign torque are most similar to one another. 
To quantify this result, Table \ref{tab:p values} shows the $p$-values from a two-sample KS test for all pairwise combinations of the spin up, spin down, and reversed torque simulations.
\begin{table*}
	\centering
	\begin{tabular}{|c|c|c|c|c|c|c|c|c|}
		& \multicolumn{2}{c}{Experiment I} & \multicolumn{2}{c}{Experiment II} &  
		\multicolumn{2}{c}{Experiment III} & \multicolumn{2}{c}{Experiment IV}   \\
		& $\abssize$ & $\deltat$ & $\abssize$ & $\deltat$ & $\abssize$ & $\deltat$ & $\abssize$ & $\deltat$  \\
		\hline
		Experiment I & \multicolumn{2}{c}{---} & $4.4 \times 10^{-3}$ & $2.9 \times 10^{-4}$ & $8.8 \times 10^{-3}$ & $6.4 \times 10^{-3} $& 0.91 & 0.94 \\
		Experiment II & \multicolumn{2}{c}{---} & \multicolumn{2}{c}{---}  & 0.72 & 0.33 & $1.4 \times 10^{-3}$  & $1.3 \times 10^{-3}$ \\
		Experiment III &  \multicolumn{2}{c}{---} & \multicolumn{2}{c}{---} & \multicolumn{2}{c}{---} & $5.9 \times 10^{-3}$ & 0.059 \\
		\hline
		&  &  &  &  
	\end{tabular}
	\caption{$p$-values from a two-sample KS test for glitch and anti-glitch sizes $\abssize$ and waiting time $\deltat$ from experiments I -- IV.
	The KS test is performed for all pairwise combinations of the four experiments. }
	\label{tab:p values}
\end{table*}
Table \ref{tab:p values} shows that the null hypothesis that the glitch populations are drawn from the same distribution is rejected with high probability for the pairs I/III and II/IV.
The null hypothesis is not rejected for the simulations with the same sign of the torque, the pairs I/IV and II/III. 
The results in Figure \ref{fig:kdes reversed} and Table \ref{tab:p values} confirm that spin up and spin down are not simply time-reversed analogues, and that the difference between the PDFs for glitches and anti-glitches is not due to the difference in $N_{\rm v}$ at the beginning or the end of the simulations.

\subsection{Avalanche locations}
\label{subsec:avalanche location}

\begin{figure}
	\begin{center}
		\includegraphics[scale=0.5]{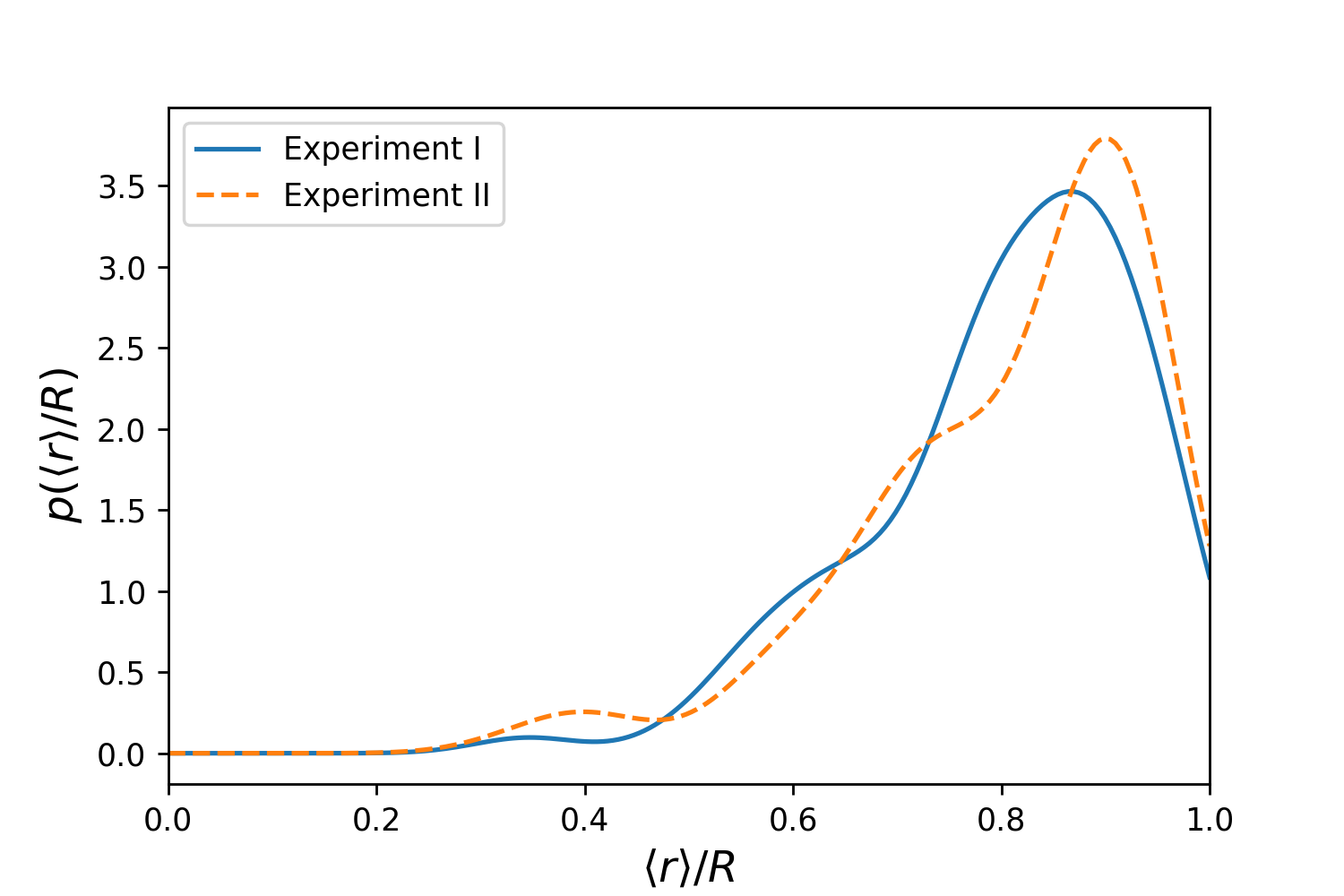}
	\end{center}
	\caption{Kernel density estimate of the PDF of mean \textbf{starting} radial coordinate $\langle r \rangle$ of vortices involved in avalanches in spin up (blue curve) and spin down (orange dashed curve).}
	\label{fig:mean radius}
\end{figure}

In spin-up simulations vortices are continually added near the boundary. 
This may lead to a bias for avalanches to begin at $r \approx R$.
To examine whether such a bias exists, for each glitch and anti-glitch in experiments I and II we identify the vortices which move $\Delta r >a$ during the avalanche.
We record the radial coordinate of each vortex in the avalanche at the beginning of the avalanche and compute the mean $\langle r \rangle$ for each avalanche.
Figure \ref{fig:mean radius} shows the PDF of $\langle r/R \rangle$ for both the spin-up and spin-down simulations.
The PDFs are similar for glitches and anti-glitches. 
This indicates that the discrepancy between the size and waiting time distributions is not due to where vortices unpin.

\subsection{Stress and spatial correlations}
\label{subsec:spatial correlations}

We now look at the distribution of stress in experiments I and II.
We parametrise the stress of a vortex as
$v_{i, \rm{stress}} = \vert \mathbf{v}_{i, \rm{induced}} - \boldsymbol{\Omega}_{\rm C} \times \mathbf{x}_i \vert / \rm{max} \vert \mathbf{v}_{\rm pin} \vert$
where $\mathbf{v}_{i, \rm{induced}}$ is the velocity contribution from the first two terms in equations \eqref{eq:x velocity} and \eqref{eq:y velocity} and $\rm{max} \vert \mathbf{v}_{\rm pin} \vert = V_0 \xi e^{-1/2}$ is the maximum velocity contribution from a Gaussian pinning site with depth $V_0$ and characteristic width $\xi$.
A vortex always unpin for $v_{i, \rm{stress}} > 1$, though it may unpin at lower values.
\citet{Howitt2020} 
found that stress is evenly distributed throughout the container and remains at $\langle v_{\rm stress} \rangle \approx 0.45$ for the duration of a simulation, and that glitches do not substantially reduce the amount of stress in the system even temporarily, e.g. the stress fluctuates by $\approx 10 \%$ in the immediate aftermath of a glitch.

\begin{figure}
	\centering
	\includegraphics[scale=0.5]{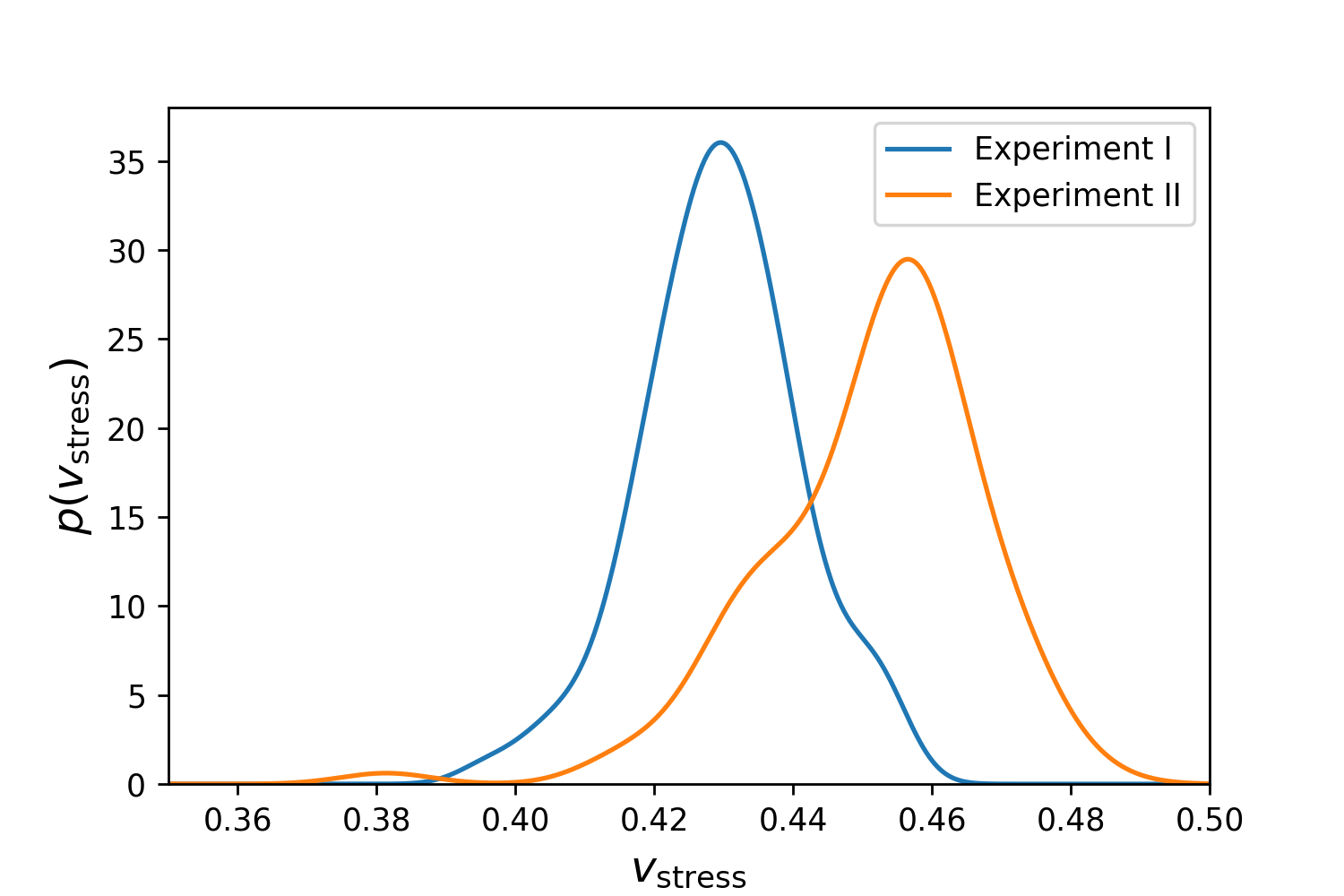}
	\caption{PDFs of vortex-averaged pinning stress $v_{\rm stress}$ in experiments I  (blue curve) and II (orange curve), sampled at glitch epochs.}
	\label{fig:mean stress}
\end{figure}

Figure \ref{fig:mean stress} shows the PDF of the vortex-averaged value of $\vstress$ at each of the glitch epochs for experiments I and II,
i.e. we compute the stress of all vortices at each glitch epoch and take the average over the array. 

Figure \ref{fig:mean stress} shows that the vortices in experiment II are under more stress prior to glitches than in experiment I. 
The mean value of $\vstress$ (averaged across the glitch epochs) is 0.43 in experiment I and 0.45 in experiment II.
This result on its own does not necessarily explain the differences between the size and waiting time PDFs in experiments I and II. 
It may in fact be the case that the smaller, more frequent avalanches in experiment I are the cause of the lower average stress rather than a consequence.

\begin{figure}
	\centering
	\includegraphics[scale=0.5]{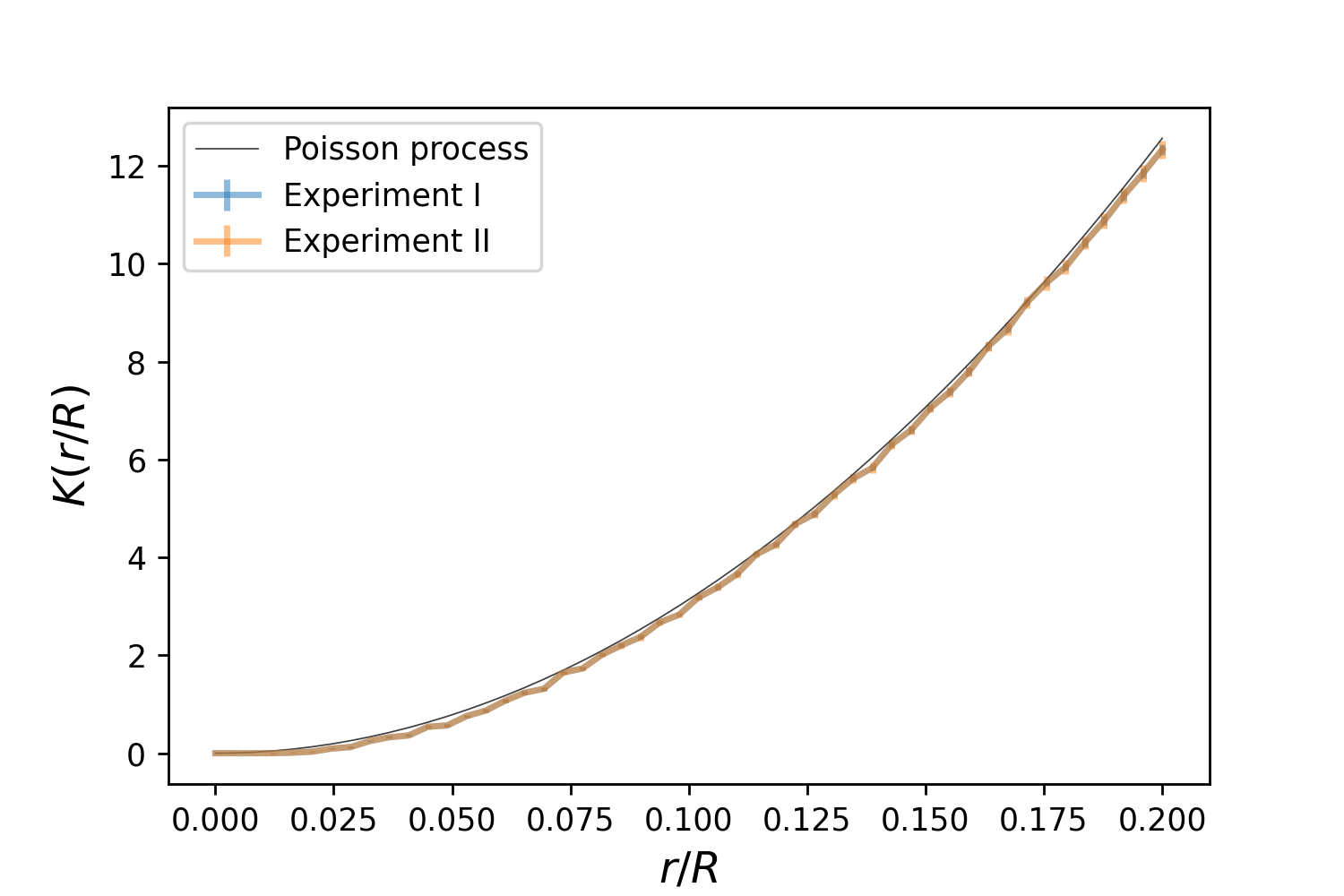}
	\caption{Ripley's $K$-function for experiment I (blue curve with error bars) and experiment II (orange curve with error bars).
		Also shown is the theoretical  value for a Poisson process $K_{\rm Poisson}(r) = \pi r^2$ (gray curve).}
	\label{fig:ripley k unmarked}
\end{figure}

We also test whether there is a difference in spatial clustering of vortices in experiments I and II using Ripley's $K$-function
\citep{Ripley1977}.
The empirical $K$-function is defined as 
\begin{equation}
	\hat{K}(r) = \frac{A}{N_{\rm v}(N_{\rm v}-1)} \sum_{i=1}^{N_{\rm v}} \sum_{j=1,j \neq i}^{N_{\rm v}}
	\mathbf{1} \lbrace r_{ij} \leq r \rbrace e_{ij} \, ,
	\label{eq:ripley k} 
\end{equation}
where $A$ is the area of the observation window, $\mathbf{1} \lbrace r_{ij} \leq r \rbrace\\$ is an indicator function that returns 1 if $r_{ij} \leq r$ and 0 otherwise, and $e_{ij}$ is an edge-correction term.
The $K$-function can be considered a measure of the amount of clustering on a length scale $r$.
In the case of a Poisson point process (i.e. an uncorrelated spatial distribution), the $K$-function can be determined analytically as $K(r) = \pi r^2$. 
A point process that has $\hat{K}(r) > \pi r^2$ implies clustering of the points on the length scale $r$, and $\hat{K}(r) < \pi r^2$ implies anti-clustering. 
Figure \ref{fig:ripley k unmarked} shows $\hat{K}(r)$ for experiments I and II. 
Equation \ref{eq:ripley k} is computed for each experiment at every glitch epoch and averaged in time with three standard deviation error bars. 
Figure \ref{fig:ripley k unmarked} shows that there is no difference between $\hat{K}(r)$ for experiments I and II, and that both are consistent with a Poisson point process.
An extension to the Ripley's $K$-function for marked point processes (such as the vortex coordinates marked by the value of $\vstress$) exists, called the mark-weighted point process
\citep{Penttinen1992}.
We also compute this on the data from experiments I and II and find nearly identical results to those shown in Figure \ref{fig:ripley k unmarked}.
We find no evidence of spatial correlations in the spatial distributions of vortices or the distributions of stress in either experiment. 

\subsection{Vortex wandering}
\label{subsec:vortex creep}

One obvious difference between experiments I and II is that in experiment I vortices are continually added to the container while in experiment II $N_{\rm v}$ remains constant except during avalanches, when some vortices move out past the container boundary.
As shown in Table \ref{tab:numerical experiments}, during experiment I the rate at which vortices are added is several times larger than the rate at which glitches occur. 
It is possible that the addition of vortices at random angular positions near $r=R$ perturbs the quasi-equilibrium of the pinned vortex array and triggers avalanches.
If this is the case, we expect that an analogous process in the spin-down case should increase the glitch rate.

To test this hypothesis, we perform two further numerical experiments where we perturb the vortex array at the same cadence that vortices are added in experiments I and IV.
In experiment V, every time $\OmegaC$ decreases by $\kappa / R^2$ we choose a vortex at random and move it in a randomly chosen direction (horizontal or vertical) by a distance equal to the pinning lattice spacing $a$.
In experiment VI, every time $\OmegaC$ decreases by $\kappa / R^2$ we find the vortex with the maximum radial position $r<R$ and move it outside the container boundary.
The wandering vortex motion in experiments V and VI is analogous though not exactly equivalent to vortex creep
\citep{Alpar1984}.

\begin{figure}
	\centering
	\includegraphics[scale=0.5]{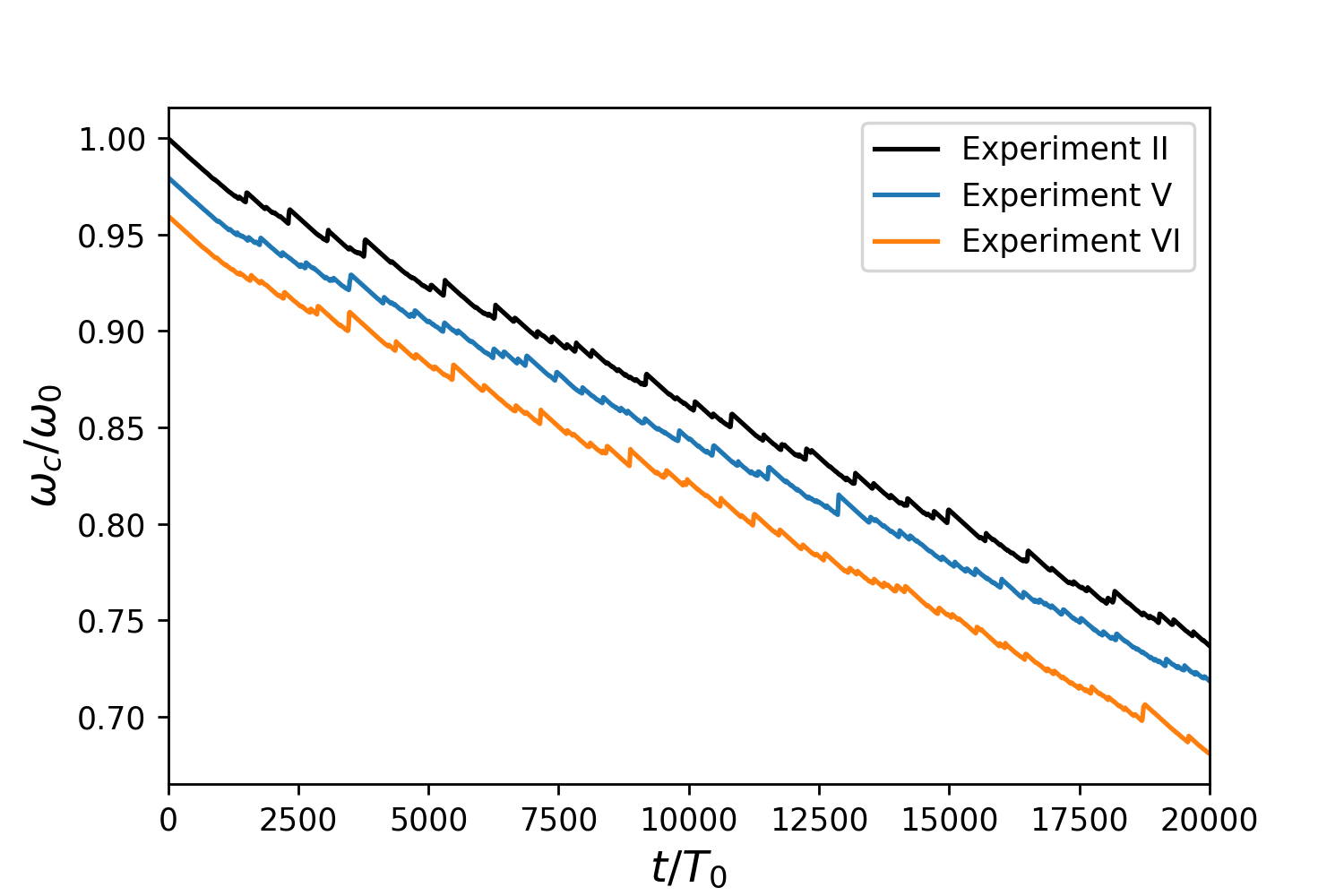}
	\caption{Container angular velocity versus time for experiment II (black curve), experiment V (blue curve) and experiment VI (orange curve).
		Results for experiments V and VI are offset vertically for clarity.}
	\label{fig:omega creep}
\end{figure}

We show the evolution of $\OmegaC$ in experiments V and VI compared to the result from experiment II in Figure \ref{fig:omega creep} and summarise the results in Table \ref{tab:creep results}.
\begin{table*}
	\centering
	\begin{tabular}{|l|l|l|l|l|}
		& Experiment I & Experiment II & Experiment V & Experiment VI \\
		\hline
		$N_{{\rm v},i}$ & 1943 & 1943 & 1943  & 1943 \\ 
		$\Omega_i/\Omega_0$ & 1 & 1 & 1 & 1 \\
		$N_{{\rm v},f}$ & 2361 & 1498 & 1492 & 1336 \\
		$\Omega_f/\Omega_0$ & 1.27 & 0.74 & 0.74 & 0.72 \\
		$N_{\rm glitches}$ & 176 & 101 & 123 & 90 \\
		\hline
	\end{tabular}
	\caption{Summary of numerical experiments I, II, V and VI, showing initial and final numbers of vortices $N_{{\rm v},i}$ and $N_{{\rm v},f}$, initial and final container angular velocities $\Omega_i$ and $\Omega_f$ and number of glitches $N_{\rm glitches}$}
	\label{tab:creep results}
\end{table*}
As shown in Table \ref{tab:creep results}, neither experiment V or VI increases $N_{\rm glitches}$ more than $\approx 30 \%$ from experiment II, compared to the $\approx 75 \%$ more glitches in experiment I.
In fact, we see fewer glitches in experiment VI than in experiment II.
A two-sample KS test for the sizes and waiting times in experiments V and VI rejects the null hypothesis that they are drawn from the same distributions as in experiment I, but not in experiment II.

While the idealized vortex wandering algorithms used in experiments V and VI are not exactly equivalent to the addition of vortices at the boundary in experiment I, the results here do suggest that the increase in the number of glitches in experiment I compared to experiment II is not simply the result of perturbations to the vortex array caused by adding vortices randomly, and the interplay of global angular momentum transfer and local pinning dynamics discussed in Section \ref{subsec:differences} remains the most plausible cause.	

\bsp	
\label{lastpage}
\end{document}